\pdfoutput=1
\documentclass[prx,twocolumn,showpacs,amsmath,amssymb,superscriptaddress,floatfix,longbibliography,nofootinbib]{revtex4-2}
\usepackage{times}
\usepackage{helvet}
\usepackage{courier}
\usepackage{hyperref}
\hypersetup{
  hypertexnames=false,
  colorlinks=true, linkcolor=cyan, citecolor=magenta, filecolor=magenta, urlcolor=cyan, runcolor=cyan
}
\usepackage{graphicx}
\usepackage{orcidlink}
\usepackage{algorithm}
\usepackage{algorithmic}
\usepackage{newfloat}
\usepackage{listings}
\usepackage{amsmath}
\usepackage{amsfonts}
\usepackage{amssymb}
\usepackage{amsthm}
\usepackage{braket}
\usepackage[capitalize]{cleveref}
\usepackage{xcolor}
\usepackage{booktabs}
\usepackage{tikz}
\usepackage{mathtools}
\usepackage{subfigure}

\newtheorem{mytheorem}{Theorem}
\crefname{mytheorem}{Theorem}{Thm}
\newtheorem{mylemma}{Lemma}
\crefname{mylemma}{Lemma}{Lem}
\newtheorem{mycorollary}{Corollary}
\crefname{mycorollary}{Corollary}{Cor}

\newcommand{\bes} {\begin{subequations}}
\newcommand{\ees} {\end{subequations}}
\newcommand{\beq} {\begin{equation}}
\newcommand{\eeq} {\end{equation}}

\def\>{\rangle}
\def\<{\langle}

\begin{document}

\title{GraphSpec: Whole-Chip Hamiltonian Spectroscopy by Graph Coloring}

\author{Amy F. Brown\enspace\orcidlink{0000-0001-6664-6494}}
\affiliation{Department of Physics \& Astronomy, University of Southern California,
Los Angeles, California 90089, USA}
\affiliation{Center for Quantum Information Science \& Technology, University of
Southern California, Los Angeles, California 90089, USA}
\author{Daniel A. Lidar\enspace\orcidlink{0000-0002-1671-1515}}
\affiliation{Department of Physics \& Astronomy, University of Southern California,
Los Angeles, California 90089, USA}
\affiliation{Center for Quantum Information Science \& Technology, University of
Southern California, Los Angeles, California 90089, USA}
\affiliation{Department of Electrical \& Computer Engineering, University of Southern California,
Los Angeles, California 90089, USA}
\affiliation{Department of Chemistry, University of Southern California, Los Angeles,
California 90089, USA}
\affiliation{Quantum Elements, 2829 Townsgate Road, Westlake Village, California 91361, USA}

\begin{abstract}
Efficient and accurate characterization of the Hamiltonian of a network of interconnected qubits is essential for calibration, control, error correction, and quantum algorithms.
As qubit counts grow, serial approaches to spectroscopy and calibration become inefficient and less reliable, which motivates their parallelization.
In this work we present Graph Spectroscopy (GraphSpec), a scalable protocol to extract all one- and two-body coefficients of a static $Z$-type Hamiltonian across a hardware graph of arbitrary connectivity, using a number of whole-chip measurement settings bounded linearly in the distance-2 chromatic number of the graph, the minimum number of colors needed to color a graph's vertices so that distinct vertices at graph distance at most two have different colors.
With ideal instantaneous pulses, modified chromatic-Hadamard dynamical decoupling (CHaDD) sequences isolate the selected coefficients at the end of every completed sequence cycle.
Using this protocol, we extract detunings and couplings on the subgraph of the 84-qubit Rigetti Ankaa-3 processor induced by its operable qubits, and we test the extracted detunings by applying the corresponding corrective drive frequency shifts.
This reduces the spread of the detuning estimates of the updated qubits by a factor of 4 in a single subsequent whole-chip GraphSpec scan.
GraphSpec thus serves calibration as well as characterization, and the extracted parameters can inform updates of device simulation models.
\end{abstract}

\maketitle

\section{Introduction}
Frequent spectroscopy is necessary to maintain the up-time and operability of online quantum computing devices: qubit frequencies~\cite{Burnett2019decoherence} and coherence times~\cite{Klimov:2018ww,Carroll2022dynamics} fluctuate, and gate calibrations built on stale values degrade accordingly~\cite{Kelly2018physical}. As qubit counts grow, the traditional serial approach of measuring one qubit at a time becomes untenable, since it prolongs the recalibration period and reduces the ratio of up-time to down-time. Beyond a certain size, recalibration becomes so time-intensive that the qubits drift substantially on the timescale of the recalibration itself. Parallelizing these measurements is therefore a prerequisite for maintaining large devices in calibration.

In addition to the intentionally engineered couplings, there are many spurious interactions~\cite{Sarovar2020detectingcrosstalk,Mundada2019,Ku2020,Kandala2021demonstration}, and these too can be measured. Accurate and rapidly refreshed knowledge of qubit frequencies and qubit-qubit interactions serves two further purposes: it tightens the loop between device characterization and design, and it supplies device parameters accurate enough to construct and study a reliable model of a given device~\cite{Wittler2021integrated,shen2026qmctwinmasterequationsimulationsyndrome} without occupying the quantum hardware itself. Accurate and scalable multi-qubit spectroscopy thus stands to benefit both characterization and simulation efforts, by increasing the rate at which device information can be obtained.

Parallel spectroscopy schemes must contend with the fact that simultaneously operated neighboring qubits shift each other's frequencies~\cite{tripathi2021suppression,evert2024syncopated}. Selective recoupling provides a way to separate these contributions by retaining chosen interactions while suppressing the rest~\cite{Wu2002recoupling}.
Applied to Ramsey spectroscopy, this principle allows the $ZZ$ coupling of a pair of qubits to be isolated by synchronized dynamical decoupling (DD) of the pair, and selective decoupling of the remaining interactions allows disjoint pairs to be measured in parallel~\cite{evert2024syncopated}.
Extending such measurements across an entire device requires a systematic choice of which terms can be probed jointly. 
This grouping problem also arises in general Hamiltonian learning, where Ref.~\cite{Huang2023learning} uses Hamiltonian reshaping by random Pauli operations to separate the dynamics into subsystems whose parameters can be learned in parallel.
For derivative estimation, Ref.~\cite{Gu2024practical} instead organizes simultaneously inferable groups of terms using a distance-2 coloring of an interaction graph, in which vertices represent Hamiltonian terms, and edges join terms with overlapping supports.

Here, we introduce a complementary parallel method we call Graph Spectroscopy (GraphSpec), a deterministic whole-chip Ramsey protocol for one- and two-body $Z$-type Hamiltonians on hardware graphs of arbitrary connectivity.
GraphSpec colors the physical-qubit graph, rather than the Hamiltonian-term graph used for derivative estimation in Ref.~\cite{Gu2024practical}, and applies the same circuit to all qubits of a given color.
We use a chromatic-Hadamard dynamical decoupling (CHaDD)~\cite{brown2024efficient} sequence modified to retain selected detunings or disjoint $ZZ$ couplings while suppressing the remaining terms. For static $Z$-type Hamiltonians with ideal instantaneous pulses, this isolation is exact at completed cycles, allowing the retained coefficients to be extracted directly from Ramsey frequencies. Scheduling these measurements by color yields an explicit bound, linear in the hardware graph's distance-2 chromatic number (the minimum number of colors needed to color a graph's vertices so that distinct vertices at graph distance at most two have different colors), on the number of whole-chip Ramsey settings needed to measure every detuning and coupling. This bound is independent of device size for graph families with bounded distance-2 chromatic number. 

We demonstrate GraphSpec on the 84-qubit Rigetti Ankaa-3 quantum processing unit (QPU) and proceed to test the extracted detunings by applying corrective drive frequencies in a single whole-chip retuning round.
These corrections reduce the spread of the detuning estimates of the updated qubits by a factor of 4, demonstrating the utility of GraphSpec for whole-chip frequency retuning with a number of Ramsey settings bounded independently of chip size for the aforementioned graph families.

The structure of this paper is as follows. In \cref{sec:background} we review the relevant background: single-qubit Ramsey spectroscopy, the graph Hamiltonian and spectator-induced effects, higher-distance graph coloring, CHaDD, and round-robin tournament scheduling. In \cref{sec:chadd-graphspec} we present our main results: CHaDD GraphSpec and the number of whole-chip experiments it requires. In \cref{sec:experiments} we present proof-of-concept characterization and detuning-correction experiments on the Rigetti Ankaa-3 QPU. In \cref{sec:beyond-z} we derive the survival rules for non-$Z$-type terms. We conclude in \cref{sec:conc} and discuss next steps. An alternative protocol, mean-field-approximation (MFA) GraphSpec, and its Ramsey derivations are presented in \cref{sec:mfa-graphspec}, with experiments in \cref{sec:mfa-graphspec-experiments} and a detuning-correction comparison between the two protocols in \cref{app:comparison}.

\section{Background}
\label{sec:background}

In this section we present the single-qubit noise model and frame conventions, introduce the graph Hamiltonian and spectator effects in parallel Ramsey spectroscopy, and review the combinatorial ingredients of GraphSpec: higher-distance graph coloring, CHaDD, and round-robin tournament scheduling.

\subsection{Single-Qubit Noise Model}
\label{sec:1q-model}

A single uncoupled qubit $v$ with frequency $\omega_v^q$
has the lab-frame Hamiltonian ($\hbar=1$ units are used throughout)
\begin{align}
\label{eq:1q-qubit-hamiltonian-lab-frame}
    H_v^\text{qubit} &= -\frac12 \omega_v^q \sigma_v^z
    ,
\end{align}
with the $-1/2$ coefficient chosen so that
$\omega_v^q = \bra{1}H_v^\text{qubit}\ket{1} - \bra{0}H_v^\text{qubit}\ket{0}$.
Single-qubit gates are implemented by a drive at frequency $\omega_v^d$, which ideally coincides with $\omega_v^q$. To work in the frame co-rotating with the drive, we move to the interaction picture generated by
\begin{align}
\label{eq:1q-interaction-picture-transformation}
    H_v^0 \equiv -\frac12 \omega_v^d \sigma_v^z
    \ , \ \
    U_v^0(t) = e^{-iH_v^0t}
    .
\end{align}
Defining the qubit detuning as
\beq
\label{eq:detuning-definition}
\Delta_v^\text{det} \equiv \omega_v^d - \omega_v^q ,
\eeq
and the qubit detuning Hamiltonian as
\bes
\begin{align}
\label{eq:1q-detuning-hamiltonian-lab-frame}
    H_v^\text{det} &\equiv H_v^\text{qubit} - H_v^0 = \frac12 \Delta_v^\text{det} \sigma_v^z
    ,\\
\label{eq:1q-detuning-hamiltonian-drive-frame}
    \tilde{H}_v^\text{det} &= {U_v^0}^\dagger H_v^\text{det} {U_v^0} = H_v^\text{det}
    ,
\end{align}
\ees
we see that the detuning term is unchanged by the transformation to the drive frame.

\subsection{$T_2^*$ Ramsey Experiments}
\label{sec:ramsey}

In the presence of detuning, but in the absence of any drive term, the $\ket{+}$ state evolves according to $\tilde{H}_v^\text{det}$ in the rotating drive frame and precesses about the $z$-axis with angular frequency $\Delta_v^\text{det}$:
\begin{align}
    \exp\left[ -i \tilde{H}_v^\text{det} t \right] \ket{+} &= \frac{e^{-i \Delta_v^\text{det} t / 2}}{\sqrt2} \left( \ket{0} + e^{i \Delta_v^\text{det} t} \ket{1} \right) .
\end{align}
The closed-system fidelity of the $\ket{+}$ state over time is then
\begin{align}
    \left| \bra{+} \exp\left[ -i \tilde{H}_v^\text{det} t \right] \ket{+} \right|^2 = \frac12+\frac12\cos\left( \Delta_v^\text{det} t \right)
    .
\end{align}
While we can determine the magnitude of the detuning $|\Delta_v^\text{det}|$ from the resulting oscillation, we cannot determine the sign, because the cosine is even.
To remedy this, we can instead read out the population of the state $\ket{+y}\equiv(\ket{0}+i\ket{1})/\sqrt2$, so that
\begin{align}
    \left| \bra{+y} \exp\left[ -i \tilde{H}_v^\text{det} t \right] \ket{+} \right|^2 = \frac12+\frac12\sin\left( \Delta_v^\text{det} t \right)
    .
\end{align}
More commonly, an artificial detuning $\omega_a$ is applied after the idle period in the form of a virtual Z rotation
\begin{align}
    R_z(\omega_a t)_v = \exp\left( -i \omega_a t \sigma_v^z / 2 \right) ,
\end{align}
which makes the closed-system fidelity of the $\ket{+}$ state
\begin{align}
\label{eq:ramsey-frequency}
    F_+ &= \left| \bra{+} R_z(\omega_a t)_v \exp\left( -i \tilde{H}_v^\text{det} t \right) \ket{+} \right|^2
    \nonumber \\
    & = \frac{1}{2}+\frac{1}{2}\cos\left[ (\omega_a+\Delta_v^\text{det}) t \right]
    .
\end{align}
By using a sufficiently large artificial detuning $\omega_a$, roughly an order of magnitude larger than the true detuning~\cite{evert2024syncopated}, we can ensure that both the magnitude and sign of $\Delta_v^\text{det}$ extracted from the resulting fringes are unambiguous.
The artificial detuning has the added benefit of increasing the rate of oscillations so that we can measure and fit several fringes even for rapidly dephasing qubits whose maximum measurement time is limited.

In a standard $T_2^*$ Ramsey experiment, we prepare a qubit in the $\ket{+}$ state (or another equator state with $\langle \sigma^z \rangle = 0$), allow the qubit to idle for a period of time, apply an artificial detuning proportional to the idle period in the form of a virtual $R_z(\varphi)$ rotation, and finally repeat the rotation used to prepare the state at the beginning of the experiment, transferring $\ket{+}$ state population to $\ket{1}$, before measuring in the $\sigma^z$ basis.
This circuit is repeated a number of times, or ``shots,'' and the proportion of $\ket{1}$ measurements is plotted as a function of time,
with a sampling rate greater than twice the highest signal frequency,
giving oscillations at $\omega_a+\Delta_v^\text{det}$ with a decay envelope characterized by $T_2^*$.
In our experiments, we modify this pre-measurement step slightly by transferring $\ket{+}$ state population to $\ket{0}$ instead of $\ket{1}$ so that we can fit to the closed-system fidelity formula in \cref{eq:ramsey-frequency}.
The exception is the detuning-correction test of \cref{sec:detuning-correction} below, in which we transfer $\ket{+y}$ state population to $\ket{0}$ instead, so that the cosine in \cref{eq:ramsey-frequency} becomes a sine and the sign of $\omega_a+\Delta_v^\text{det}$ can be read off from the phase of the fringes.

\subsection{Graph Hamiltonian and Noise Model}
\label{sec:graph-hamiltonian-noise-model}

Consider a network of interconnected qubits modeled by a weighted graph $G=(V,E)$ with qubit frequencies $\omega_v^q$ as vertex weights and qubit-qubit coupling frequencies $J_{uv}$ as edge weights corresponding to the lab-frame graph Hamiltonian
\bes
\label{eq:graph-hamiltonian}
\begin{align}
\label{eq:graph-hamiltonian-components}
    H_G &= H^\text{qubit} + H^\text{int}, \\
\label{eq:qubit-hamiltonian}
    H^\text{qubit} &= \sum_{v \in V} H_v^\text{qubit}, \\
\label{eq:interaction-hamiltonian}
    H^\text{int} &= \sum_{uv \in E} H_{uv}^\text{int} \ , \ \
    H_{uv}^\text{int} = \frac14 J_{uv} \sigma_u^z \sigma_v^z, 
\end{align}
\ees
with $H_v^\text{qubit}$ as given in \cref{eq:1q-qubit-hamiltonian-lab-frame}.
We transform to the interaction picture defined by each qubit's drive frequency $\omega_v^d$ via
\begin{align}
\label{eq:interaction-picture-transformation}
    H_0 &= \sum_{v \in V} H_v^0 \ , \ \
    U_0(t) = e^{-iH_0t} = \prod_{v \in V} U_v^0(t) ,
\end{align}
with $H_v^0$ and $U_v^0(t)$ as given in \cref{eq:1q-interaction-picture-transformation}.
The qubit detuning Hamiltonian is
\begin{align}
\label{eq:detuning-hamiltonian-lab-frame}
    H^\text{det} &\equiv H^\text{qubit} - H_0 = \sum_{v \in V} H_v^\text{det}
\end{align}
with $H_v^\text{det}$ as given in \cref{eq:1q-detuning-hamiltonian-lab-frame},
and the detuning and interaction Hamiltonians transform as
\bes
\begin{align}
\label{eq:detuning-hamiltonian-drive-frame}
    \tilde{H}^\text{det} &= U_0^\dagger H^\text{det} U_0 = H^\text{det}
    ,\\
\label{eq:interaction-hamiltonian-drive-frame}
    \tilde{H}^\text{int} &= U_0^\dagger H^\text{int} U_0 = H^\text{int}
    .
\end{align}
\ees
The total lab-frame graph Hamiltonian then transforms to the drive frame as
\bes
\begin{align}
\label{eq:lab-frame-graph-hamiltonian}
    H_G &= H_0 + H^\text{det} + H^\text{int}
    ,\\
\label{eq:drive-frame-graph-hamiltonian}
    \tilde{H}_G
    &=
    \sum_{v \in V} \frac12 \Delta_v^\text{det} \sigma_v^z
    +
    \sum_{uv \in E} \frac14 J_{uv} \sigma_u^z \sigma_v^z
    .
\end{align}
\ees
The goal of GraphSpec is to determine every vertex weight $\Delta_v^\text{det}$ and every edge weight $J_{uv}$ of the drive-frame graph Hamiltonian $\tilde{H}_G$ in as few whole-chip Ramsey measurements as possible.

Under \cref{eq:drive-frame-graph-hamiltonian}, the Ramsey signal of a measured qubit depends on the states of its coupled neighbors, called spectators~\cite{tripathi2021suppression}. When all spectators occupy $\sigma^z$ eigenstates with eigenvalues $s_u=\pm1$, qubit $v$ precesses at $\Delta_v^\text{det}+\frac12\sum_{u:uv\in E}J_{uv}s_u$. Preparing spectators on the equator instead of in $\sigma^z$ eigenstates introduces modulation of the Ramsey envelope. We derive these effects in \cref{sec:mfa-graphspec}, where they provide the basis for MFA GraphSpec.

\subsection{Higher-Distance Graph Coloring}
\label{sec:graph-coloring}

Graph coloring is an extensively studied problem in discrete mathematics, and it underlies the efficient multi-qubit dynamical decoupling provided by the CHaDD protocol~\cite{brown2024efficient}, which we review in \cref{sec:chadd} and adapt to GraphSpec in \cref{sec:chadd-graphspec} below.

Let $d(u,v)$ denote the graph distance between vertices $u$ and $v$ (the number of edges in a shortest path connecting them), and define the closed neighborhood of a vertex $v$ as
\beq
\label{eq:N_v}
N_v \equiv \{ u \in V ~|~ d(u,v) \leq 1 \}.
\eeq

A $C$-coloring of a graph $G=(V,E)$ is a function
$f:V \rightarrow \{0,\dots,C-1\}$
that assigns colors $c=0,\dots,C-1$ to the vertices of the graph.
This coloring is proper if no edge connects vertices of the same color, i.e., $uv \in E \implies f(u) \neq f(v)$.
The smallest number of colors that can be used to properly color the graph is the chromatic number, $\chi$, so any proper $C$-coloring must satisfy $C\geq\chi$.
Note that any graph coloring $f$ partitions the vertex and edge sets as follows:
\bes
\begin{align}
\label{eq:vertex-set-color-partition}
    V_c &= \{ v \in V ~|~ f(v) = c \} \\
\label{eq:edge-set-color-partition}
    E_{c_1,c_2} &= \{ uv \in E ~|~ f(u)=c_1, f(v)=c_2 \}
\end{align}
\ees
and that a proper coloring enforces that there be no edge with $c_1=c_2$, i.e., $E_{c,c} = \emptyset$.

\begin{figure*}
\centering
(a)
\includegraphics[width=0.98\columnwidth]{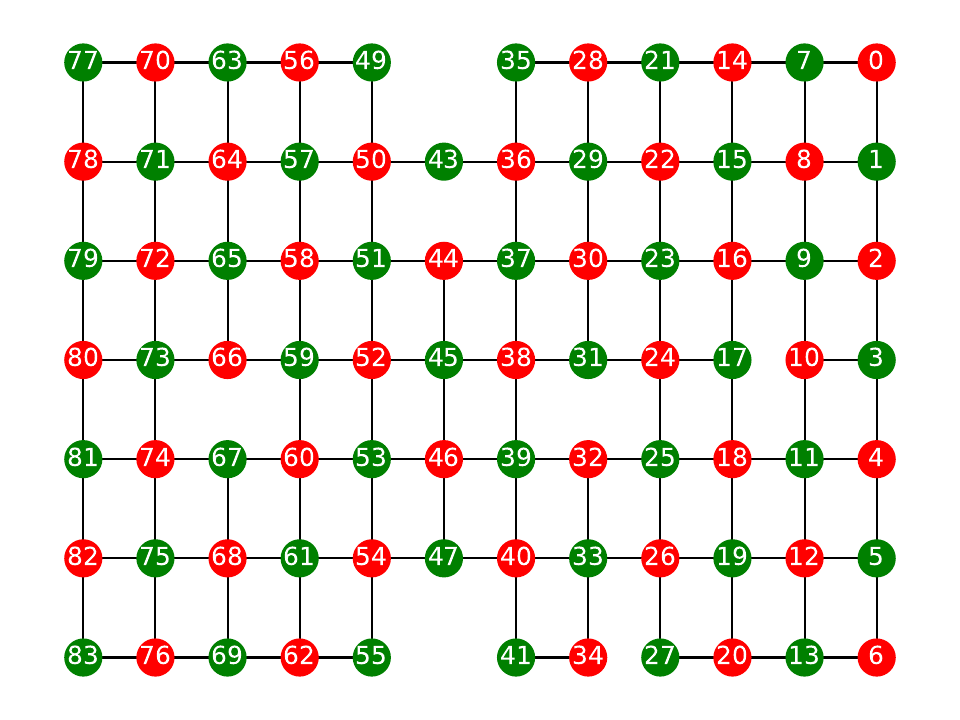}
(b)
\includegraphics[width=0.98\columnwidth]{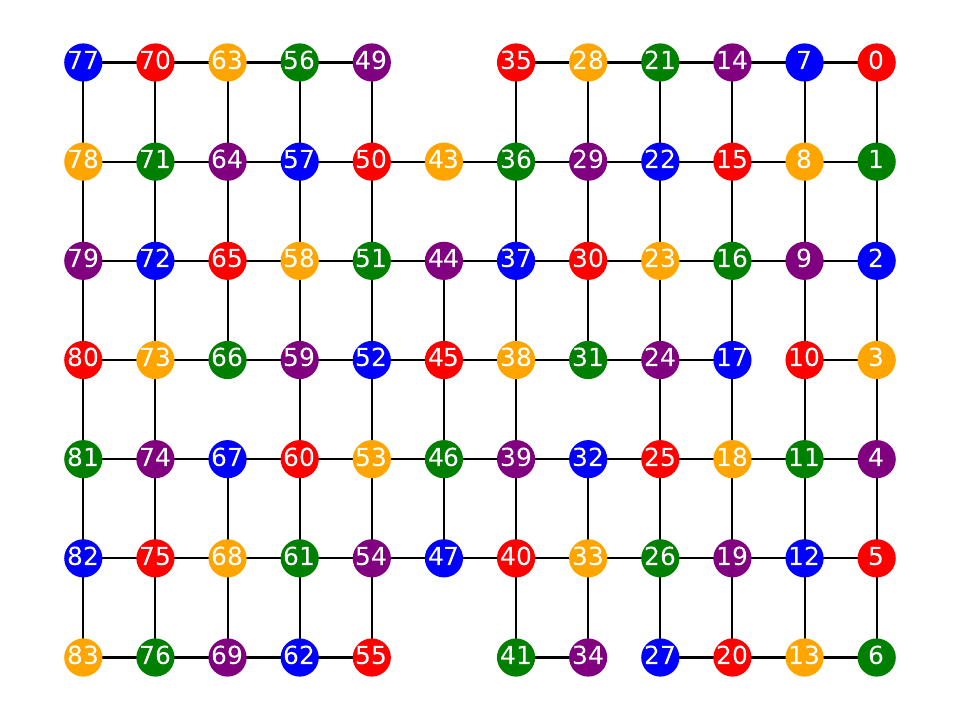}
\caption{
(a) (Distance-1) 2-coloring and
(b) distance-2 5-coloring of Rigetti Ankaa-3.
Qubits 42 and 48 are inoperable and missing from the graph;
their detunings and incident couplings are excluded from the reported parameter extraction.
Some edges are missing, e.g., (43, 44), meaning that the corresponding two-qubit gate is unavailable;
however, the qubit-qubit coupling exists and can be measured.
}
\label{fig:ankaa3-distance2-5coloring}
\end{figure*}

Higher-distance graph coloring is a more constrained version of standard, i.e., distance-1, graph coloring.
A distance-$d$ coloring requires that any two vertices at most distance $d$ apart on the graph be colored differently, i.e., $d(u,v)\leq d \implies f(u) \neq f(v)$.
Equivalently, the closest any two qubits of the same color can be is $d+1$, i.e., $f(u) = f(v) \implies d(u,v) > d$, with both implications restricted to $u \neq v$.
The smallest number of colors for which this can be accomplished is the distance-$d$ chromatic number, $\chi_d$.
An example of a distance-1 2-coloring and a distance-2 5-coloring of the Rigetti Ankaa-3 square lattice appear in \cref{fig:ankaa3-distance2-5coloring}(a) and (b), respectively.

The following two lemmas are elementary consequences of the definition of a distance-$d$ coloring and are well known~\cite{Gebremedhin2005what}; we include proofs for completeness, in the form used by our theorems below.
\begin{mylemma}
\label{lem:disjoint-neighborhoods}
Let $f$ be a proper distance-$d$ coloring of $G$ with $d \geq 2$.
Then the closed neighborhoods of any two distinct vertices of the same color are disjoint, i.e., $N_u \cap N_v = \emptyset$ for all $u \neq v \in V_c$ and every color $c$, where $N_v$ is the closed neighborhood of a vertex $v$ defined in \cref{eq:N_v}.
\end{mylemma}
\begin{proof}
Let $u \neq v \in V_c$, so that $f(u)=f(v)$ and hence $d(u,v) > d \geq 2$.
Suppose, for contradiction, that there exists a vertex $w \in N_u \cap N_v$.
Then $d(u,w) \leq 1$ and $d(w,v) \leq 1$, so the triangle inequality gives $d(u,v) \leq d(u,w) + d(w,v) \leq 2$, contradicting $d(u,v) > 2$.
\end{proof}

\begin{mylemma}
\label{lem:neighborhood-distinct-colors}
Let $f$ be a proper distance-$d$ coloring of $G$ with $d \geq 2$.
Then any two distinct vertices in the same closed neighborhood have distinct colors, i.e., $f(u) \neq f(w)$ for all $u \neq w \in N_v$ and every $v \in V$.
In particular, every vertex has at most one neighbor of any given color.
\end{mylemma}
\begin{proof}
Let $u \neq w \in N_v$, so that $d(u,v) \leq 1$ and $d(v,w) \leq 1$.
By the triangle inequality, $d(u,w) \leq d(u,v) + d(v,w) \leq 2 \leq d$.
Since $f$ is a proper distance-$d$ coloring, it follows that $f(u) \neq f(w)$.
\end{proof}

The distance-1 2-coloring in \cref{fig:ankaa3-distance2-5coloring}(a) shows that $d=1$ suffices for neither lemma: it permits two vertices of the same color at distance 2, whose closed neighborhoods then intersect, and it permits a vertex with two neighbors of the same color.

\begin{mycorollary}
    \label{cor:square-lattice-chromatic-numbers}
    For the square lattice, and for any subgraph containing a vertex of degree four, $\chi=2$ and $\chi_2=5$.
\end{mycorollary}
\begin{proof}
Let $v=(x,y)$ denote a square-lattice vertex with integer coordinates, and define the color assignment functions $f_1(v)\equiv(x+y)\bmod2$ and $f_2(v)\equiv(x+2y)\bmod5$.
Nearest-neighbor displacements change $f_1$ by $1$ modulo $2$, so $f_1$ is a proper distance-1 coloring.
For any nonzero displacement $(\delta x,\delta y)$ with $|\delta x|+|\delta y|\leq2$, the change $\delta x+2\delta y$ belongs to $\{\pm1,\pm2,\pm3,\pm4\}$ and is therefore nonzero modulo $5$, so $f_2$ is a proper distance-2 coloring.
Restricting either function to a subgraph preserves the corresponding coloring property.
An edge requires two colors, and the closed neighborhood of a degree-four vertex requires five distinct colors by \cref{lem:neighborhood-distinct-colors}.
These upper and lower bounds coincide for the Ankaa-3 graph, which contains degree-four vertices.
\end{proof}

\subsection{Chromatic-Hadamard Dynamical Decoupling (CHaDD)}
\label{sec:chadd}

CHaDD~\cite{brown2024efficient} combines graph coloring and multi-qubit dynamical decoupling based on orthogonal arrays~\cite{Stollsteimer:01,Rotteler:2006aa}
to decouple the Hamiltonian of an arbitrary hardware graph in a number of time steps that scales linearly with the number of colors in the proper coloring of the graph, i.e., $N \in \mathcal{O}(C)$.

An $N \times N$ Hadamard matrix with $N=2^\nu$,
\begin{align}
\label{eq:hadamard-matrix}
    \bra{i} W_\nu \ket{j}
    =
    (-1)^{i \cdot j}
    ,
\end{align}
where $i \cdot j$ denotes the mod-2 inner product of the binary representations of $i$ and $j$, satisfies $W_\nu W_\nu^T = N I$.
The inner product of any two distinct rows vanishes, since for $i_1 \neq i_2$ exactly half of the values of $j$ satisfy $(i_1 \oplus i_2) \cdot j = 1$, where $\oplus$ denotes bitwise addition mod 2.
In particular, the sum of the elements of any row $i \neq 0$, equal to its inner product with row $0$, also vanishes:
\bes
\begin{align}
\label{eq:hadamard-rows-orthogonal}
    \sum_{j=0}^{N-1} (-1)^{(i_1 \oplus i_2) \cdot j}
    &=
    N \delta_{i_1 i_2}
    ,\\
\label{eq:hadamard-rows-sum-to-zero}
    \sum_{j=0}^{N-1} (-1)^{i \cdot j}
    &=
    N \delta_{i,0}
    .
\end{align}
\ees
Each color $c=0,\dots,C-1$ is mapped to a row of the Hadamard matrix via a color-to-row map $g:\mathbb{Z}_C\rightarrow\mathbb{Z}_N$, and the CHaDD decoupling operators $U_j$ that conjugate the Hamiltonian in each time step $j=0,\dots,N-1$ are
\begin{align}
\label{eq:chadd-decoupling-operators}
    U_j
    =
    \bigotimes_{c=0}^{C-1} \widetilde{X}_c^{g(c) \cdot j}
    ,\hspace{4mm}
    \widetilde{X}_c \equiv \bigotimes_{v \in V_c} X_v
    ,
\end{align}
where $X_v$ denotes a $\pi$ rotation about $x$, equal to $\sigma_v^x$ up to an irrelevant phase.
The operator $U_j$ is the cumulative control during interval $j$.
Writing $U_{-1}=I$, the boundary pulse is $U_jU_{j-1}^\dagger$; a final pulse $U_{N-1}^\dagger$ closes the cycle after the $N$ equal intervals of duration $\tau$.

The CHaDD statements below assume ideal instantaneous pulses and equal free-evolution intervals of duration $\tau$.
The proof of the single-axis CHaDD theorem in Ref.~\cite{brown2024efficient} gives the following result for an arbitrary proper coloring.

\begin{mytheorem}[Single-axis CHaDD]
\label{thm:chadd-cgdd}
Let $f$ be a proper $C$-coloring of $G$, and choose $N=2^{\lfloor\log_2 C\rfloor+1}$.
If $g:\mathbb{Z}_C\rightarrow\mathbb{Z}_N$ is injective and $0\notin\mathrm{Im}(g)$, the CHaDD sequence of \cref{eq:chadd-decoupling-operators} has circuit depth $C<N\leq2C$ and cancels, to first order in $\tau$ at fixed $N$, every time-independent one- and two-qubit Pauli term supported on $G$, except single-qubit $X$ and two-qubit $XX$ terms.
\end{mytheorem}

Ref.~\cite{brown2024efficient} states the result for a minimum coloring, but its proof uses only $f(u)\neq f(v)$ whenever $uv\in E$.
The chosen $N$ supplies at least $C$ nonzero Hadamard rows, so we can assign a distinct row to every color without determining the chromatic number.
A term containing one $Y$ or $Z$ factor acquires the corresponding row's sign and averages to zero by \cref{eq:hadamard-rows-sum-to-zero}.
A two-qubit term containing two such factors acquires the product of the rows assigned to its endpoint colors; these rows are distinct because $f$ is proper and $g$ is injective, so the term averages to zero by \cref{eq:hadamard-rows-orthogonal}.
The single-qubit $X$ and two-qubit $XX$ terms commute with every pulse and remain.

For the static $Z$-type Hamiltonian of \cref{eq:drive-frame-graph-hamiltonian}, these remaining terms are absent.
GraphSpec instead requires control over which terms survive, which we obtain by allowing an arbitrary $g$.
The following variant characterizes the surviving terms and is the premise of CHaDD GraphSpec.

\begin{mytheorem}[CHaDD with an arbitrary color-to-row map]
\label{thm:chadd}
Let $f$ be a proper $C$-coloring of $G$, let $g:\mathbb{Z}_C \rightarrow \mathbb{Z}_N$ be an arbitrary color-to-row map with $N = 2^\nu$, and let the corresponding CHaDD sequence be given by \cref{eq:chadd-decoupling-operators}.
For the static $Z$-type Hamiltonian of \cref{eq:drive-frame-graph-hamiltonian}, the completed cycle retains exactly the detuning terms of the colors assigned to row $i=0$ and the coupling terms of the edges whose two endpoint colors are assigned to a common row:
\bes
\label{eq:average-hamiltonians}
\begin{align}
\label{eq:average-detuning-hamiltonian}
    \overline{H}^\text{det}
    &=
    \sum_{c=0}^{C-1}
    \delta_{g(c),0}
    \sum_{v \in V_c} \frac12 \Delta_v^\text{det} \sigma_v^z
    ,\\
\label{eq:average-interaction-hamiltonian}
    \overline{H}^\text{int}
    &=
    \sum_{c_1 < c_2}
    \delta_{g(c_1),g(c_2)}
    \sum_{uv \in E_{c_1,c_2}} \frac14 J_{uv} \sigma_u^z \sigma_v^z
    .
\end{align}
\ees
\end{mytheorem}

\begin{proof}
Each decoupling operator [\cref{eq:chadd-decoupling-operators}] conjugates the $Z$-type terms according to
\beq
\label{eq:z-conjugation}
U_j^\dagger \sigma_v^z U_j = (-1)^{g(f(v)) \cdot j} \sigma_v^z ,
\eeq
so in time step $j$ a detuning term of color $c$ acquires the sign $(-1)^{g(c) \cdot j}$, and a coupling term on an edge in $E_{c_1,c_2}$ acquires the sign $(-1)^{g(c_1) \cdot j} (-1)^{g(c_2) \cdot j} = (-1)^{[g(c_1) \oplus g(c_2)] \cdot j}$.
Averaging over the $N$ time steps and using \cref{eq:hadamard-rows-sum-to-zero} and \cref{eq:hadamard-rows-orthogonal}, respectively, yields \cref{eq:average-detuning-hamiltonian,eq:average-interaction-hamiltonian}.
In \cref{eq:average-interaction-hamiltonian}, the restriction $c_1<c_2$ counts each edge exactly once: proper coloring excludes edges within a color class, and $E_{c_1,c_2}=E_{c_2,c_1}$ because edges are unordered pairs.
All $U_j^\dagger\tilde{H}_G U_j$ are diagonal in the computational basis and therefore commute. After the closing pulse, the cycle propagator is consequently
\beq
\mathcal{U}(T)=\exp\left[-iT\left(\overline{H}^\text{det}+\overline{H}^\text{int}\right)\right],
\qquad T=N\tau,
\eeq
which proves exact isolation at completed cycles.
In particular, if no color is assigned to row $i=0$, i.e., $0$ is not in the image of $g$, and $g$ is injective, then every term vanishes and \cref{thm:chadd-cgdd} is recovered for $Z$-type Hamiltonians: assigning each of the $C$ colors to a distinct row other than row $i=0$ requires $N \geq C+1$, i.e., a Hadamard matrix $W_\nu$ with $\nu = \lfloor\log_2 C\rfloor+1$, so the circuit depth is $C < N = 2^{\lfloor\log_2 C\rfloor+1} \leq 2C$.
\end{proof}

For Hamiltonians with noncommuting toggled terms, exponentiating the cycle-averaged Hamiltonian generally gives only the leading-order cycle evolution~\cite{brown2024efficient}; we return to this limitation in \cref{sec:beyond-z} below.

To measure every coupling, we must schedule color pairs so that each pair shares a row once, while no color belongs to two pairs in the same round. A round-robin tournament supplies this schedule.

\subsection{Round-Robin Tournament Scheduling}
\label{sec:round-robin}

A single round-robin tournament schedules one-on-one games among $n$ players such that every player plays every other player exactly once, in as few rounds $R$ as possible.
There are ${n \choose 2}=\frac{n(n-1)}2$ combinations of two players, so this is the total number of games to be played.
If $n$ is even (odd), up to $P=\frac{n}2$ ($\frac{n-1}2$) games can be played in each round
(with one player sitting out each round if $n$ is odd),
and if the maximum number of games is played at a time,
all games can be played in $R=n-1$ ($n$) rounds.
To give every player at least one round off, a ``dummy'' player can be added to the schedule; the dummy's opponent has a ``bye,'' i.e., sits out that round.

A single round-robin tournament schedule can be represented by the $R \times P$ matrix $S$ whose entry in row $r=0,\dots,R-1$ and column $p=0,\dots,P-1$ consists of an ordered pair of players.
By construction, every combination of two players appears somewhere in $S$, and if a dummy is included, every player is paired with it (and hence sits out) at least once.

An example of a single round-robin tournament schedule for $n=5$ players, in which every player plays every other player once and has at least one bye, is given by the matrix $S$, which schedules players $0,1,\dots,4$ and a dummy player $5$:
\begin{align}
\label{eq:ex-c5-round-robin-schedule}
    S
    =
    \begin{bmatrix}
        (0,5) & (1,4) & (2,3) \\
        (0,4) & (5,3) & (1,2) \\
        (0,3) & (4,2) & (5,1) \\
        (0,2) & (3,1) & (4,5) \\
        (0,1) & (2,5) & (3,4)
    \end{bmatrix}
    ,
\end{align}
where the tuple $\bra{r}S\ket{p}$ in row (round) $r$ and column (pair slot) $p$ lists the two players matched in that game.
This schedule has the two properties we need: each combination of two players appears exactly once, and each player has exactly one bye (a game against the dummy player $5$).

\section{CHaDD GraphSpec}
\label{sec:chadd-graphspec}

We now construct the color-to-row maps needed to measure every coefficient of \cref{eq:drive-frame-graph-hamiltonian}.
We count a round as one whole-chip preparation, pulse-pattern, and readout setting, sampled at several evolution times and over repeated shots. The number of settings or rounds is $R$, whereas $N$ counts the time steps in each CHaDD cycle. 

We remark that the bounds we derive below on $R$ do not specify the shots or total evolution time needed at a given precision. In addition,
the coefficient extraction statements below assume resolved Ramsey frequencies whose signs are fixed as in \cref{sec:ramsey}, with sampling fast enough to avoid aliasing.

As stated at the end of \cref{sec:chadd}, permitting colors to be assigned to row $i=0$, or relaxing the injectivity of $g$, repurposes CHaDD to decouple all but a chosen subset of the terms of $\tilde{H}_G$.
By \cref{eq:average-detuning-hamiltonian}, the detuning terms of qubits of any color assigned to row $i=0$ survive the CHaDD average; by \cref{eq:average-interaction-hamiltonian}, so do the coupling terms of edges connecting qubits of colors assigned to the same row.
All other terms average to zero; by \cref{thm:chadd}, the surviving terms generate the exact evolution at integer multiples of the cycle duration.

In CHaDD GraphSpec, we apply the Ramsey preparation and readout of \cref{sec:ramsey} in parallel and replace the idle period by a CHaDD sequence, so that a qubit on the equator evolves stroboscopically under the averaged Hamiltonian $\overline{H}_G=\overline{H}^\text{det}+\overline{H}^\text{int}$, defined in \cref{eq:average-hamiltonians}.

\begin{mytheorem}[CHaDD GraphSpec to extract detunings and couplings]
\label{thm:chadd-graphspec-detunings-couplings}
$R \leq C+1$ rounds of CHaDD GraphSpec,
with CHaDD sequence circuit depth $P \leq N < 2P$,
where $P$ is the number of color pairs per round,
are sufficient to determine all detunings $\Delta_v^\text{det} ~\forall~ v \in V$ and couplings $J_{uv}~\forall~uv \in E$ in \cref{eq:drive-frame-graph-hamiltonian}, represented by a graph with distance-2 chromatic number $\chi_2 \leq C$.
\end{mytheorem}
\begin{proof}
Let $f$ be a proper distance-2 $C$-coloring of $G=(V,E)$, and let $S$ be a single round-robin tournament schedule for the $C$ real colors together with one (two) ``dummy'' colors if $C$ is odd (even), so that the total number of players is even.
The schedule then has $R = C$ ($R = C+1$) rounds with $P = (C+1)/2$ ($P = C/2 + 1$) pairs per round, so $R \leq C+1$.

In each round $r=0,\dots,R-1$, schedule a CHaDD sequence by assigning the two colors of each pair of real colors to a common nonzero row of the Hadamard matrix, with distinct pairs assigned to distinct rows.
Assign the color paired with the dummy (the first dummy, if there are two) to row $i=0$, and, if $C$ is even, assign the color paired with the second dummy to a nonzero row not used by any pair.
(In the round in which the two dummies are paired with each other, if any, no color is assigned to row $i=0$.)
At most $P$ distinct rows are used per round, so a Hadamard matrix with $P \leq N < 2P$ suffices.

Consider the averaged Hamiltonian $\overline{H}_G$ of a given round.
By \cref{eq:average-detuning-hamiltonian,eq:average-interaction-hamiltonian}, the only detuning terms that survive are those of the color assigned to row $i=0$, and the only coupling terms that survive are those of edges joining the two colors of a common pair; every other term averages to zero.
By \cref{lem:neighborhood-distinct-colors}, every qubit has at most one neighbor of the color with which its own color is paired, so each qubit retains at most one term: a qubit of the row-0 color retains exactly its detuning term and no coupling, while a qubit of a paired color retains at most one coupling, to its sole neighbor of the partner color.

In each pair, we designate the qubits of one color as main qubits, prepared in $\ket{+}$, and those of the other color as spectators, prepared in $\ket{0}$; the qubits of the row-0 color are also main qubits.
The states of all remaining colors are irrelevant, since their couplings to the main qubits are decoupled.
Since each spectator satisfies $\sigma_u^z\ket{0}_u=\ket{0}_u$, a retained coupling $\tfrac14 J_{uv}\sigma_u^z\sigma_v^z$ acts on main qubit $v$ as $\tfrac14 J_{uv}\sigma_v^z$.
A whole-chip Ramsey experiment then yields, on each main qubit $v$, fringes at frequency $\omega_a + \Delta_v^\text{det}$ if $v$ is of the row-0 color, or $\omega_a + \frac12 J_{uv}$ if $v$ is of a paired color with a partner-color neighbor $u$, by \cref{eq:ramsey-frequency}; a paired-color qubit with no partner-color neighbor yields fringes at $\omega_a$.

Every real color is paired with the (first) dummy exactly once, so every detuning is measured.
Every pair of real colors appears exactly once, and every edge joins two distinct colors since $f$ is proper, so every coupling $J_{uv}$, $uv \in E$, survives in exactly one round and is measured on its main-color endpoint.
Thus all detunings and couplings are determined in $R \leq C+1$ rounds.
\end{proof}

When the qubit-qubit couplings are much smaller than the qubit detunings, it may be advantageous to perform the detuning and coupling spectroscopy separately, with different artificial detunings and timescales.
The next two theorems address the two tasks separately.

\begin{mytheorem}[CHaDD GraphSpec to extract detunings]
\label{thm:chadd-graphspec-detunings}
    $R=C$ rounds of CHaDD GraphSpec measurements,
    with CHaDD sequence circuit depth $C \leq N < 2C$,
    are sufficient to determine all detunings
    $\Delta_v^\text{det} ~\forall~ v \in V$ in \cref{eq:drive-frame-graph-hamiltonian}, represented by a graph $G=(V,E)$
    with chromatic number $\chi \leq C$.
\end{mytheorem}
\begin{proof}
Let $f$ be a proper $C$-coloring of $G$.
In each round $r=0,1,\dots,R-1$, define the $r$th round color-to-row map by $i = g_r(c) \equiv c+r~\text{mod}~C$.
Every $g_r$ is injective, and since $f$ is proper, every edge of $G$ joins two distinct colors and hence two colors assigned to distinct rows; thus $\overline{H}^\text{int}=0$ in every round, by \cref{eq:average-interaction-hamiltonian}.
Note that $0 \in \mathrm{Im}(g_r)$: in round $r$, precisely the color $c \equiv -r ~\text{mod}~ C$ is assigned to row $i=0$, so by \cref{eq:average-detuning-hamiltonian} the only terms that survive are the detunings of that color:
\begin{align}
    \overline{H}_G
    =
    \sum_{v \in V_{g_r^{-1}(0)}}
    \frac12 \Delta_v^\text{det} \sigma_v^z
    .
\end{align}
Since all couplings are decoupled, applying the Ramsey preparation and readout of \cref{sec:ramsey} to all qubits, with the CHaDD sequence in place of the idle period, yields fringes at frequency $\omega_a + \Delta_v^\text{det}$ on every qubit $v$ of the row-0 color, and at $\omega_a$ on every other qubit, independently of the spectator states.

The coloring $f$ partitions the vertex set [\cref{eq:vertex-set-color-partition}]: the $V_c$ are disjoint and their union is $V$.
As $r$ runs over $0,\dots,C-1$, every color is assigned to row $i=0$ exactly once, so all detunings are obtained in $R=C$ rounds.
Since $g_r$ is injective and assigns exactly one color to row $i=0$ in each round, $N \geq C$ rows are needed; the smallest Hadamard matrix with $N \geq C$ satisfies $C \leq N < 2C$, which is the circuit depth of the CHaDD sequences replacing the idle evolution.
\end{proof}

\begin{mytheorem}[CHaDD GraphSpec to extract couplings]
\label{thm:chadd-graphspec-couplings}
    For $C \geq 2$, $R \leq C$ rounds of CHaDD GraphSpec measurements,
    with CHaDD sequence circuit depth $P < N \leq 2P$,
    where $P$ is the number of color pairs per round,
    are sufficient to determine all qubit-qubit couplings $J_{uv}~\forall~uv \in E$ in \cref{eq:drive-frame-graph-hamiltonian}, represented by a graph $G=(V,E)$ with distance-2 chromatic number $\chi_2 \leq C$.
\end{mytheorem}
\begin{proof}
Let $f$ be a proper distance-2 $C$-coloring of $G=(V,E)$, and let $S$ be a single round-robin tournament schedule for the $C$ colors, with $R = C-1$ rounds and $P = C/2$ pairs per round if $C$ is even, and $R = C$ rounds, each with one bye color, and $P = (C-1)/2$ pairs per round if $C$ is odd; in either case $R \leq C$.
In each round, schedule a CHaDD sequence by assigning the two colors of each pair to a common nonzero row of the Hadamard matrix, with distinct pairs assigned to distinct rows, and, if $C$ is odd, assign the bye color to row $i=0$.
By \cref{eq:average-interaction-hamiltonian}, the only coupling terms that survive are those of edges joining the two colors of a common pair, and by \cref{eq:average-detuning-hamiltonian}, the only detuning terms that survive are those of the bye color, which are not needed in this round (they provide a free consistency check with \cref{thm:chadd-graphspec-detunings}).
By \cref{lem:neighborhood-distinct-colors}, each qubit of a paired color therefore retains at most one coupling term: the one to its sole neighbor of the partner color, if that neighbor exists.

In each pair, we designate the qubits of one color as the ``main'' qubits, prepared in $\ket{+}$, and those of the other color as the ``spectator'' qubits, prepared in $\ket{0}$.
A whole-chip Ramsey experiment with the CHaDD sequence in lieu of the idle period then yields, on each main qubit $v$ with a partner-color neighbor $u$, fringes at frequency $\omega_a + \frac12 J_{uv}$, by \cref{eq:ramsey-frequency} and the spectator eigenvalue $+1$, and fringes at $\omega_a$ on every main qubit without such a neighbor.

Every pair of colors appears exactly once in the schedule, and every edge joins two distinct colors since $f$ is proper, so every coupling $J_{uv}~\forall~uv \in E$ is retained in exactly one round and measured on its main-color endpoint.
At most $P$ distinct nonzero rows are used per round, so a Hadamard matrix with $N \geq P+1$ rows, i.e., $P < N \leq 2P$, suffices.
\end{proof}

\begin{figure*}
\centering
\includegraphics[width=0.98\textwidth]{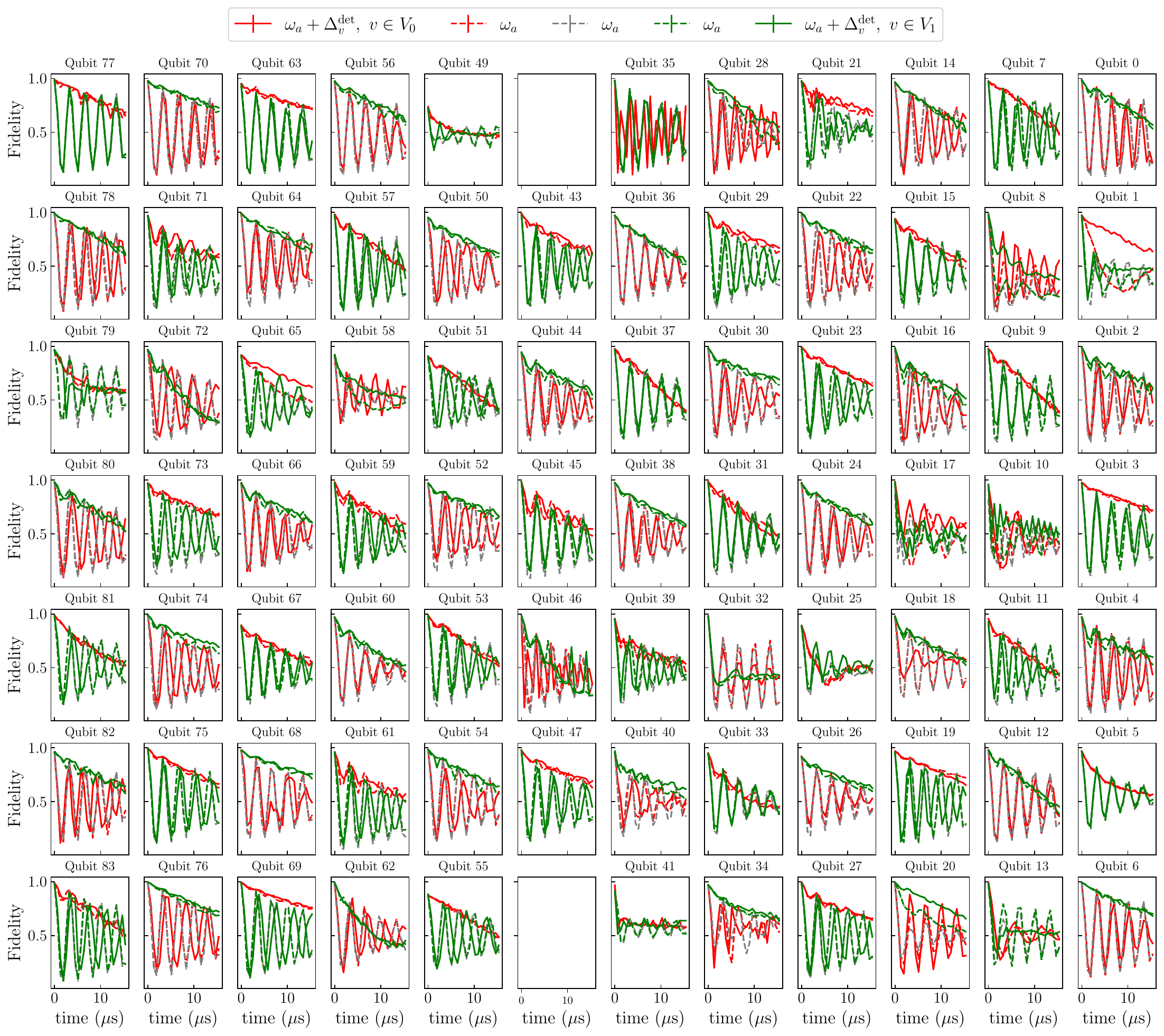}
\caption{
Five whole-chip Ramsey experiments on Ankaa-3, with one panel per qubit.
The distance-1 coloring assigns each qubit to a fixed class: red ($V_0$) or green ($V_1$).
Curve colors label the experimental preparation: red curves correspond to preparing $V_0$ in $\ket{+}$ and $V_1$ in $\ket{0}$; green curves to preparing $V_1$ in $\ket{+}$ and $V_0$ in $\ket{0}$; the gray dashed curve corresponds to preparing all qubits in $\ket{+}$.
Thus, in a red qubit's panel, the red curves show Ramsey signals and the green curves show spectator populations; these roles reverse in a green qubit's panel.
The gray curve shows a Ramsey signal in every panel.
For the Ramsey signals, solid curves use CHaDD GraphSpec to retain the measured qubit's detuning, giving the expected fringe frequency $\omega_a+\Delta_v^{\text{det}}$.
Dashed curves use plain CHaDD to cancel all detunings and couplings, giving the expected fringe frequency $\omega_a$.
}
\label{fig:chadd-graphspec-detuning-whole-chip}
\end{figure*}

\begin{figure*}
\centering
(a)
\includegraphics[width=0.98\columnwidth]{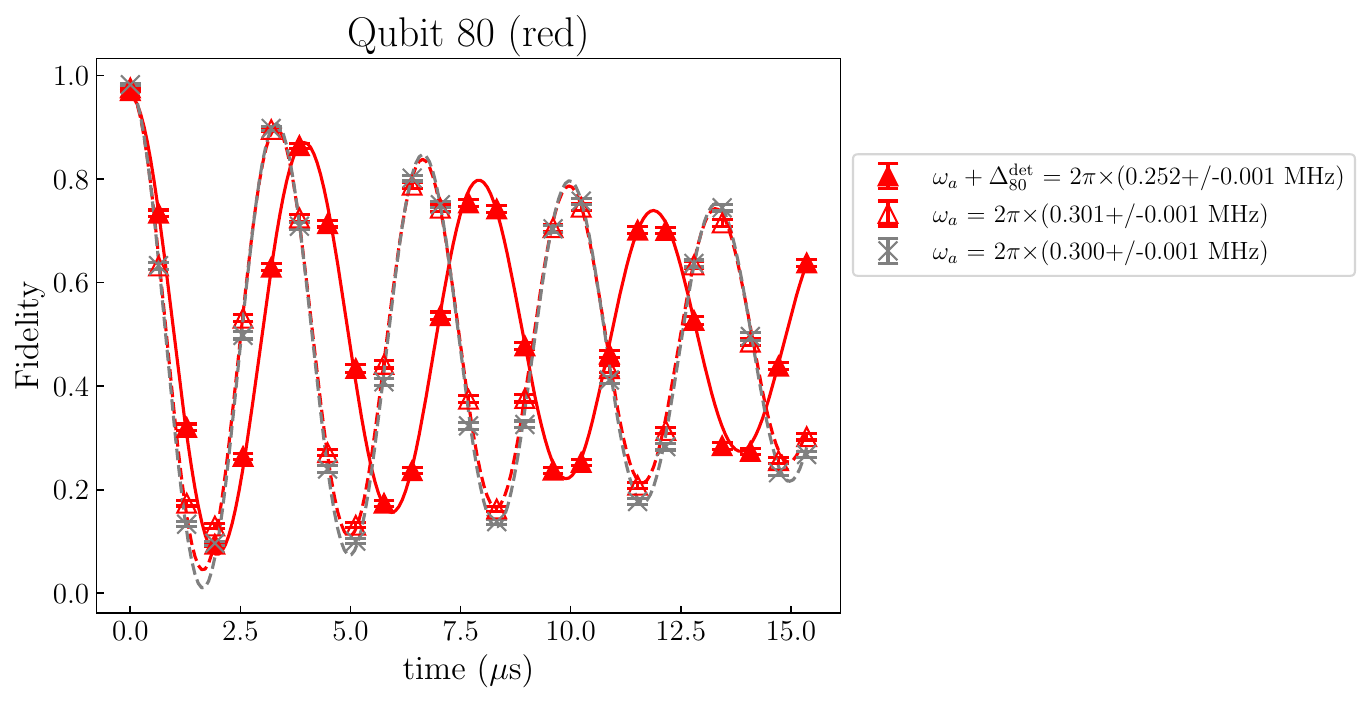}
(b)
\includegraphics[width=0.98\columnwidth]{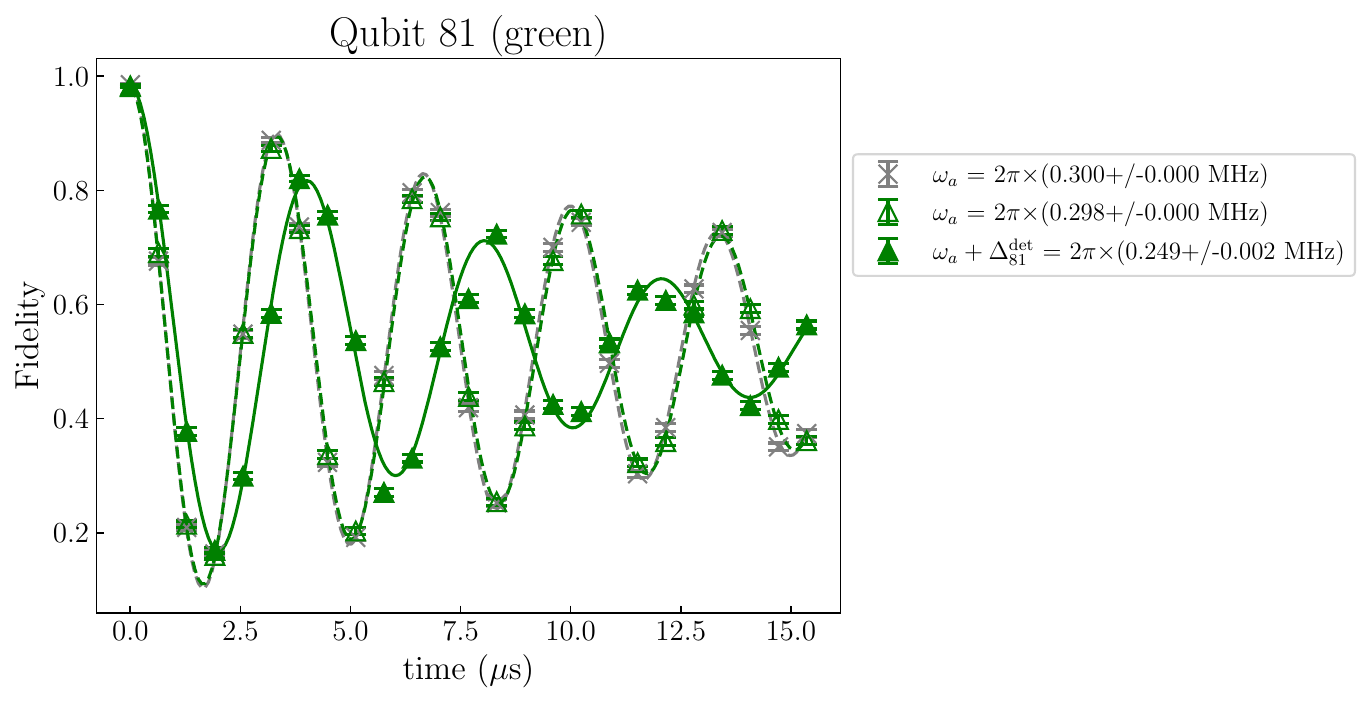}
\\
(c)
\includegraphics[width=0.98\columnwidth]{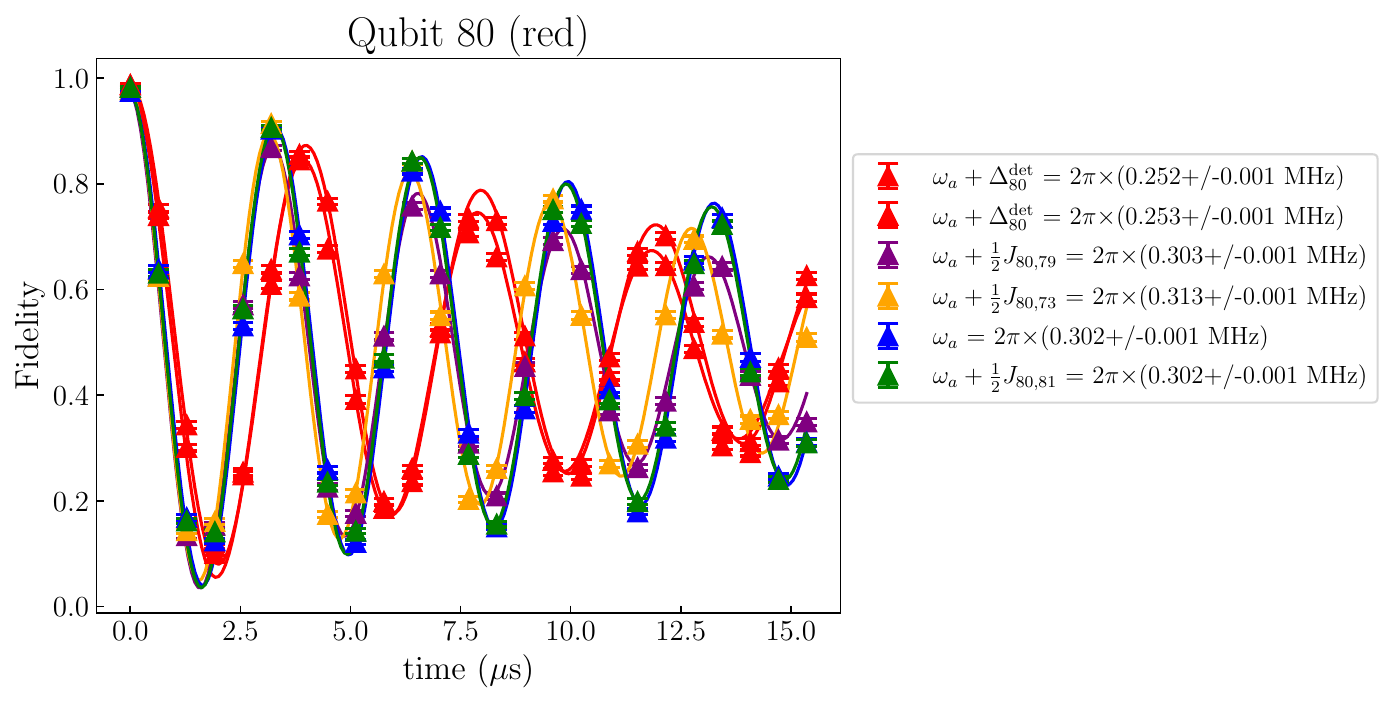}
(d)
\includegraphics[width=0.98\columnwidth]{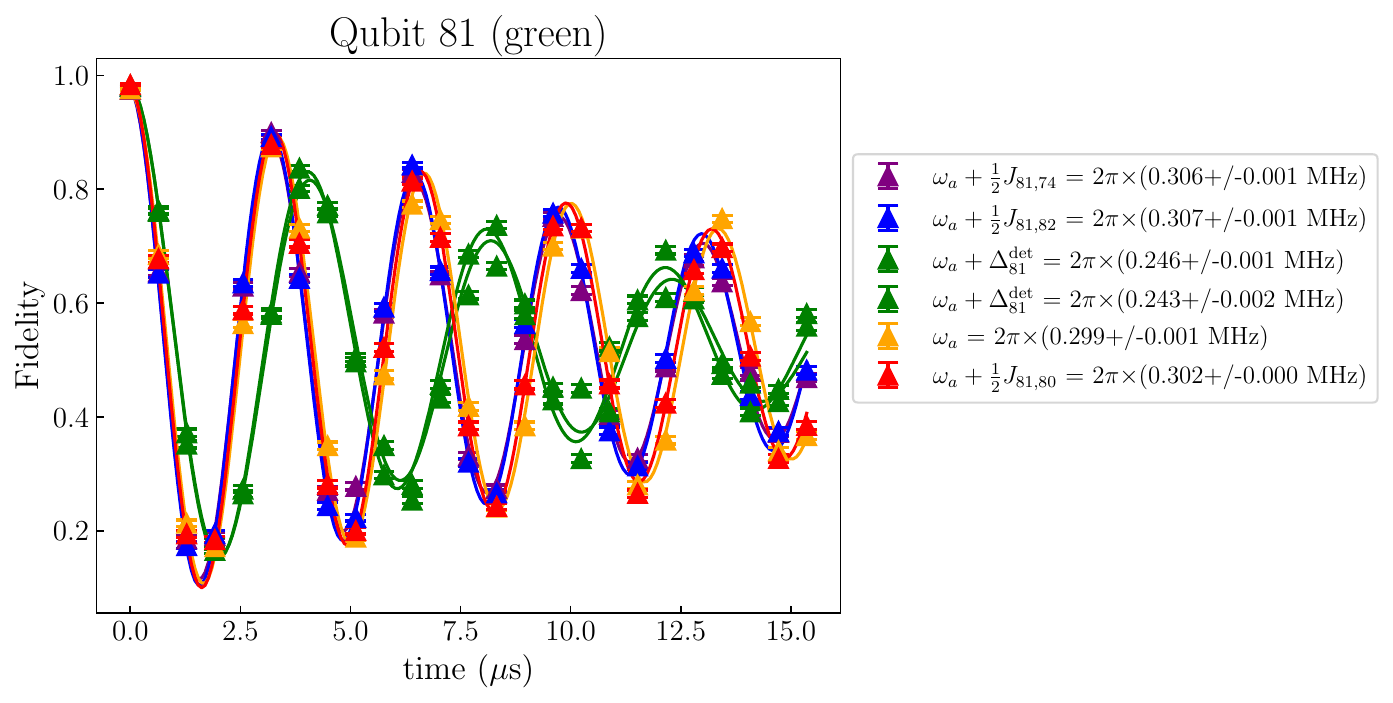}
\caption{
CHaDD GraphSpec to extract detunings (a,b) and detunings and couplings (c,d).
Qubits 80 (a,c) and 81 (b,d) are red and green, respectively, in both the distance-1 and distance-2 colorings of Ankaa-3 (\cref{fig:ankaa3-distance2-5coloring}).
Each legend lists the fitted fringe frequencies and the combination of detunings and couplings to which they correspond.
}
\label{fig:chadd-graphspec}
\end{figure*}

\section{Experiments}
\label{sec:experiments}

We performed GraphSpec experiments on the Rigetti Ankaa-3 QPU, an 84-qubit chip arranged in a $7\times12$ square grid, whose distance-1 and distance-2 colorings are shown in \cref{fig:ankaa3-distance2-5coloring}.
The depicted graph omits the inoperable qubits 42 and 48 and five edges, (10, 17), (27, 34), (31, 32), (43, 44), and (66, 67), whose corresponding two-qubit gates are unavailable (the displayed colorings remain proper when these five edges are restored to the interaction graph $G$).
An unpulsed (i.e., freely evolving) inoperable neighboring qubit can shift a measured detuning, while CHaDD measurements of couplings between operable qubits cancel the corresponding boundary interactions under the static $Z$-type model and ideal pulses.
We derive these effects and distinguish unavailable gates from absent physical interactions in \cref{app:inoperable-qubits}.
Rigetti Quantum Cloud Services provides access to Ankaa-3 by reservation rather than through a queue, so that within a reservation the time between consecutive experiments is known and no other jobs intervene.

\subsection{Procedure}
We assumed at the outset that the bare qubit detunings would be roughly $30$~kHz across the chip and the couplings roughly an order of magnitude smaller.
We therefore chose an artificial detuning of $\omega_a = 2\pi \times 300$~kHz for all our experiments, an order of magnitude larger than the anticipated detunings~\cite{evert2024syncopated}, which was chosen to keep the anticipated carrier frequencies positive and to fit several oscillation periods within the measurement window.
(Here and below, frequencies quoted in Hz refer to $\omega/2\pi$.)

When the idle period of a Ramsey experiment is replaced by a DD sequence, the signal can only be sampled at multiples of the sequence duration, and care must be taken that this sampling does not alias the underlying Ramsey signal.
All qubits on Ankaa-3 have single-qubit gates of duration $\tau_p=40$~ns, and enforcing a minimum interpulse spacing of $\tau_d=2\tau_p=80$~ns from the start of one pulse to the start of the next to reduce pulse-interference effects~\cite{vezvaee2024virtualzgatessymmetric} gives a four-step CHaDD sequence of duration $T_\text{CHaDD}=4\tau_d=320$~ns.
For an anticipated signal frequency of roughly $300$~kHz, the sequence can be repeated four times between samples while remaining below the Nyquist sampling interval, since $4T_\text{CHaDD}=1280~\text{ns} < 1/(600~\text{kHz}) \approx 1667$~ns.
Each GraphSpec experiment was sampled at 25 evenly spaced times at intervals of $640~\text{ns} = 2T_\text{CHaDD}$,
out to a maximum time of $T_\text{max}=15.36~\mu\text{s}$;
the corresponding Nyquist frequency is $1/(2\times640~\text{ns})\approx0.781$~MHz.
Each circuit was run 5000 times, and for each qubit, we computed the sample mean $\bar{F}$ of its ground-state outcomes and the standard error of the mean $\epsilon_F$.
We plot $\bar{F}$ against the sequence duration $t$, with error bars of $\pm\epsilon_F$.
We then fit $\bar{F}(t)$ to the decaying cosine model
\begin{align}
    \label{eq:decay-cosine-fit-model}
    F(t) = A\, e^{-t/T_2^*} \cos\left( \omega t + \phi \right) + B ,
\end{align}
minimizing $\chi^2$ with weights $1/\epsilon_F^2$, from the initial guesses $A=0.5$, $B=0.5$, $T_2^*=2~\mu\text{s}$, $\omega=\omega_a=2\pi\times0.3$~MHz, and $\phi=0$ (or $\phi=-\pi/2$ for the $\ket{+y}$ readout of \cref{sec:detuning-correction} below) for the amplitude, baseline, decay time, frequency, and phase, respectively.
We imposed no parameter bounds other than a positive amplitude and frequency.
The reported parameter uncertainties are the square roots of the diagonal entries of the estimated covariance matrix rescaled by the reduced $\chi^2$.

\subsection{CHaDD GraphSpec Experiments}
\label{sec:chadd-graphspec-experiments}

We performed CHaDD GraphSpec experiments on Ankaa-3 for detunings only (\cref{thm:chadd-graphspec-detunings}) and detunings and couplings (\cref{thm:chadd-graphspec-detunings-couplings}).
These require distance-1 and -2 colorings, respectively, of Ankaa-3, depicted in \cref{fig:ankaa3-distance2-5coloring}(a,b).
We ran the two sets of experiments within an hour of one another; the data for the former appear in \cref{fig:chadd-graphspec-detuning-whole-chip} and \cref{fig:chadd-graphspec}(a,b), and the data for the latter in \cref{fig:chadd-graphspec}(c,d).

By \cref{thm:chadd-graphspec-detunings}, only $R=\chi=2$ rounds of CHaDD GraphSpec are needed to extract all detunings, one for each color, red and green, of \cref{fig:ankaa3-distance2-5coloring}(a); these are the solid-line experiments in \cref{fig:chadd-graphspec-detuning-whole-chip} (whole chip) and \cref{fig:chadd-graphspec}(a,b) (qubits 80 and 81).
For comparison, we also included plain CHaDD experiments (dashed lines): a CHaDD sequence as originally designed, with injective $g$ and no color assigned to row $i=0$, decouples all detunings and couplings, leaving only the artificial detuning measurable.
We specify pulse sequences in the notation $c$: $p_0 p_1 p_2 p_3$, where every qubit of color $c$ receives pulse $p_j \in \{\texttt{I}, \texttt{X}\}$ at the end of time step $j$;
the toggling frame of color $c$ during step $j$ is then the product $p_{j-1} \cdots p_0$ (the identity for $j=0$), which realizes a row of the Hadamard matrix: \texttt{IIII}, \texttt{XXXX}, \texttt{IXIX}, and \texttt{XIXI} correspond to rows $i=0$, $1$, $2$, and $3$ of $W_2$, respectively.
The sequence used in the plain CHaDD experiments (dashed lines) is \texttt{0:~IXIX, 1:~XIXI}, which decouples all detunings and couplings of a 2-colored graph.
In the gray dashed experiment, all qubits are prepared and read out in the $\ket{+}$ state, and every qubit exhibits Ramsey fringes at the artificial detuning $\omega_a$ alone.
In the red (green) dashed experiment, all red (green) qubits are prepared in the $\ket{+}$ state and act as main qubits, while all green (red) qubits are prepared in the $\ket{0}$ state and act as spectators, and each main qubit $v$ exhibits Ramsey signals with carrier frequency $\omega_a$.
In the red (green) solid experiment, the pulse train of the red (green) qubits is replaced by the identity, i.e., that color is reassigned to row $i=0$, which retains the detuning terms of that color while still decoupling all couplings.
Again, the red (green) qubits are main qubits prepared in the $\ket{+}$ state and the green (red) qubits are spectators prepared in $\ket{0}$, and each main qubit $v$ exhibits Ramsey signals with carrier frequency $\omega_a + \Delta_v^\text{det}$.

Five rounds suffice to measure all detunings and couplings on Ankaa-3 with CHaDD GraphSpec [\cref{thm:chadd-graphspec-detunings-couplings} with $C = \chi_2 = 5$; see \cref{fig:ankaa3-distance2-5coloring}(b)], one per round of the round-robin schedule.
Using the single round-robin schedule for $C=5$ colors of \cref{eq:ex-c5-round-robin-schedule}, with $R=5$ rounds of $P=3$ pairs each,
we used the following five CHaDD GraphSpec pulse patterns:
\\\\
\begin{minipage}{\columnwidth}
\begin{verbatim}
{0: IIII, (1, 4): XIXI, (2, 3): XXXX}
{(0, 4): IXIX, 3: IIII, (1, 2): XXXX}
{(0, 3): IXIX, (4, 2): XIXI, 1: IIII}
{(0, 2): IXIX, (3, 1): XIXI, 4: IIII}
{(0, 1): IXIX, 2: IIII, (3, 4): XXXX}
\end{verbatim}
\end{minipage}
\\

Each pattern decouples the detunings of all colors except the idling one, and the couplings between all pairs of colors except those assigned to a common pulse train, whose coupling terms are left invariant.
A distance-2 coloring guarantees that each closed neighborhood consists of distinct colors (\cref{lem:neighborhood-distinct-colors}), so in each round, every qubit is either idling, in which case it is decoupled from all of its neighbors and its Ramsey fringes measure its detuning, or it is decoupled from all but its sole neighbor (if any) of the color with which its color is paired.
In each pair, one color serves as the main color, prepared in the $\ket{+}$ state, and the other as the spectator color, prepared in $\ket{0}$.
We ran two versions of each round: one in which the first color of each pair, together with the idling color, served as main (colors 0, 1, and 2 in the first round), and one in which the second color of each pair served as main.
In either case, the idling color is treated as main so that its detunings are measured in the round in which it is unpaired.
A single version per round suffices for the count in \cref{thm:chadd-graphspec-detunings-couplings}, since each coupling is measured on the main-color endpoint of its edge; the second version measures each coupling from the other endpoint and provides a consistency check.
The data for qubits 80 and 81 for the experiments in which they are main qubits are plotted in \cref{fig:chadd-graphspec}(c,d).

\subsection{Detuning Correction}
\label{sec:detuning-correction}
Ankaa-3 allows users to change the qubit drive frequencies, so we were able to test the extracted detunings directly by applying updated drive frequencies and checking that the measured detunings are strongly suppressed once they have been accounted for.
This is a self-consistency test: because the diagnostic and post-correction scans use the same estimator, an unchanged additive offset in that estimator cancels in the comparison.
The couplings, however, cannot be modified by an Ankaa-3 user, so a corresponding test of the coupling measurements was not possible.

\begin{figure}
\centering
(a)
\includegraphics[width=0.98\columnwidth]{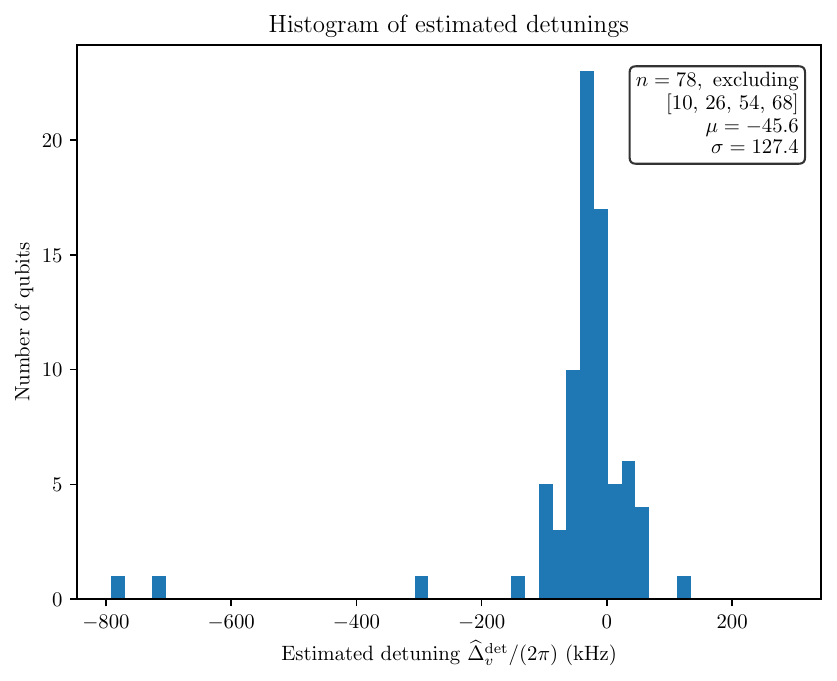}
\\
(b)
\includegraphics[width=0.98\columnwidth]{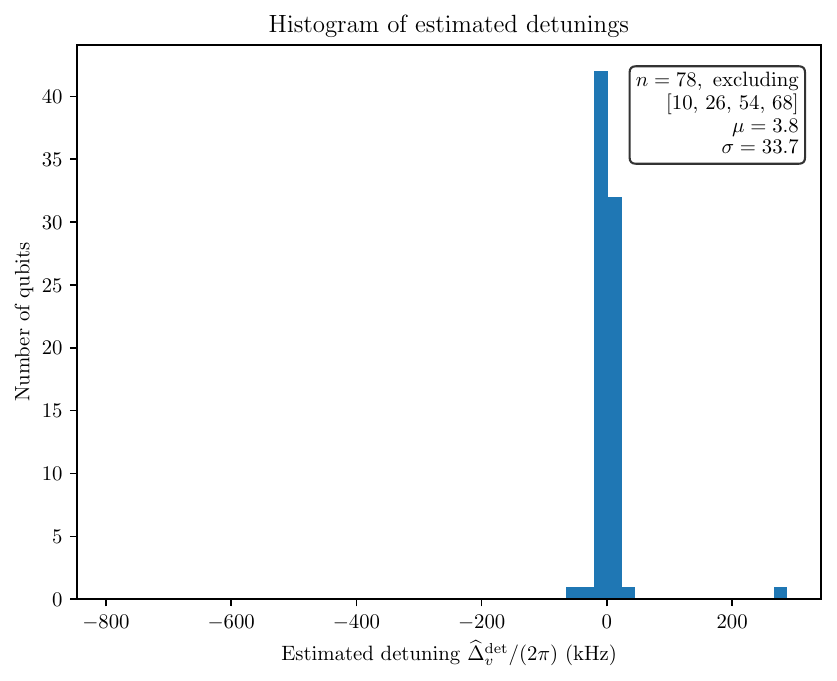}
\caption{
Detuning correction with CHaDD GraphSpec for detunings only.
Histograms of the measured qubit detunings across the chip for (a) the diagnostic scan and (b) the post-correction scan,
demonstrating a considerable reduction in both the spread and the center of the measured detunings.
Both histograms use the same bins and contain the $n=78$ qubits whose drive frequencies were updated; the annotations list the excluded qubits and give the sample mean $\mu$ and sample standard deviation $\sigma$ in kHz.
}
\label{fig:chadd-detuning-correction}
\end{figure}

We ran diagnostic and post-correction scans of CHaDD GraphSpec for detunings only within an hour of one another to minimize drift, with each scan comprising the two color-resolved settings.
We ran these scans about a week after the CHaDD GraphSpec experiments of \cref{sec:chadd-graphspec-experiments}, and read out the population of $\ket{+y}$ rather than $\ket{+}$ (\cref{sec:ramsey}).
The Ramsey signal is then $\frac12+\frac12\sin[(\omega_a+\Delta_v^\text{det})t]$, so that, with $A>0$ and $\omega>0$ in \cref{eq:decay-cosine-fit-model}, the fitted phase $\hat\phi$ distinguishes the two signs of $\omega_a+\Delta_v^\text{det}$: the detuning estimate is $\widehat{\Delta}_v^\text{det}=-\hat\omega-\omega_a$ if $\hat\phi$ lies within $\pi/10$ of $\pi/2$, and $\widehat{\Delta}_v^\text{det}=\hat\omega-\omega_a$ otherwise, where $\hat\omega$ is the fitted frequency.
From the diagnostic estimate $\widehat{\Delta}_v^\text{det}$ [\cref{fig:chadd-detuning-correction}(a)], we set
\beq
\label{eq:frequency-update}
\omega_v^{d,\mathrm{new}}=\omega_v^{d,\mathrm{old}}-\widehat{\Delta}_v^\text{det},
\eeq
consistent with \cref{eq:detuning-definition}, and repeated the scan [\cref{fig:chadd-detuning-correction}(b)].
The sample mean and sample standard deviation are annotated on each histogram.
The distribution of measured detunings narrows after the correction, and its mean moves closer to zero:
the standard deviation decreases by a factor of 3.8 (from 127.4 to 33.7~kHz), and the mean moves from $-45.6$ to $3.8$~kHz.
A few qubits dominate both standard deviations.
In \cref{fig:chadd-detuning-correction}(a) these are qubits 41, 46, and 49, the only ones whose fitted phase lies within $\pi/10$ of $\pi/2$, with estimates of $-300$, $-723$, and $-792$~kHz, respectively.
After the update the post-correction scan finds qubits 46 and 49 at $2.3$ and $-1.5$~kHz, respectively, although the fringes of qubit 49 decay within a few microseconds.
The Ramsey signal from which the detuning of qubit 41 is extracted shows almost no fringes in either scan, as in its panels in \cref{fig:chadd-graphspec-detuning-whole-chip,fig:ankaa3-graph-spec-all-qubits}: its diagnostic fit returns $\hat\omega\approx0$, i.e., the estimate $-\omega_a$, so that neither of its estimates is meaningful.
In \cref{fig:chadd-detuning-correction}(b) it is qubit 39, at $288$~kHz, whose update was in error for the reason we give below.
Without qubits 41, 46, and 49 the diagnostic standard deviation is 42.1~kHz, and without qubit 39 the post-correction standard deviation is 8.7~kHz.

The drive-frequency updates applied at the time of the experiment were computed with an earlier version of the fitting routine, which differed from the one we describe after \cref{eq:decay-cosine-fit-model} in its initial guesses, parameter bounds, and weights; the estimates plotted in \cref{fig:chadd-detuning-correction} are computed with the latter.
For every qubit plotted in \cref{fig:chadd-detuning-correction}(a) except qubit 39, the applied update $\omega_v^{d,\mathrm{old}}-\omega_v^{d,\mathrm{new}}$ agrees with the diagnostic estimate $\widehat{\Delta}_v^{\mathrm{det}}$ shown there to within 2~kHz, well below the post-correction spread.
For qubit 39 the earlier routine returned $-299$~kHz, whereas the present one returns $-5.9$~kHz; the applied update therefore overshot by $293$~kHz, and the post-correction scan finds this qubit at $288$~kHz.
Qubits 10, 26, 54, and 68 are excluded from both histograms in \cref{fig:chadd-detuning-correction} because their drive frequencies were not updated between the diagnostic and post-correction scans:
in the fits made at the time of the experiment, those of qubits 26 and 54 failed, and those of qubits 10 and 68 returned a phase more than $\pi/10$ away from its initial guess of $-\pi/2$.
Under the present routine the fits of qubits 26 and 54 succeed, and since these two qubits were not updated, their estimates give a check of reproducibility between the two scans: they change from $-9.4$ to $-16.3$~kHz for qubit 26, and from $-51.5$ to $-50.6$~kHz for qubit 54.

In the static $Z$-type model of \cref{eq:drive-frame-graph-hamiltonian}, the drive update changes the detuning contribution to the $Z$ rotation angle accumulated during an idle interval $t$ from $\Delta_v^\text{det}t$ to $(\Delta_v^\text{det}-\widehat{\Delta}_v^\text{det})t$.
CHaDD GraphSpec can therefore serve as a low-level method for correcting coherent frequency errors through calibration, with the remaining detuning-induced phase error determined by the error in the detuning estimate and any subsequent frequency drift.
\Cref{app:comparison} reports the corresponding MFA GraphSpec test and compares the two.

\section{Beyond $Z$-Type Hamiltonians}
\label{sec:beyond-z}
The theory of \cref{sec:background,sec:chadd-graphspec} assumes the purely $Z$-type Hamiltonian of \cref{eq:drive-frame-graph-hamiltonian}, but all QPUs contain non-$Z$ terms as well. Ankaa-3 uses an $XY$ interaction for its native entangling gates~\cite{Sundar2026plasma}, as do related transmon gate implementations~\cite{Abrams2020implementation,Caldwell2018parametrically}. Virtual transitions through higher transmon levels can also generate static $ZZ$ interactions~\cite{Mundada2019,Ku2020}.
In this section we characterize which terms of a general one- and two-local Hamiltonian survive a CHaDD sequence with an arbitrary color-to-row map, generalizing \cref{thm:chadd}.

Consider the general one- and two-local Hamiltonian with time-independent coefficients in the frame in which the control pulses are defined
\begin{align}
\label{eq:general-graph-hamiltonian}
    H_G^\text{gen}
    =
    \sum_{v \in V} \sum_{a} \frac12 h_v^a \sigma_v^a
    +
    \sum_{uv \in E} \sum_{a,b} \frac14 J_{uv}^{ab} \sigma_u^a \sigma_v^b
    ,
\end{align}
where $a, b \in \{x, y, z\}$, which reduces to \cref{eq:drive-frame-graph-hamiltonian} when $h_v^z = \Delta_v^\text{det}$, $J_{uv}^{zz} = J_{uv}$, and all other coefficients vanish.
Since conjugation by $X$ preserves $\sigma^x$ and flips the sign of $\sigma^y$ and $\sigma^z$, the conjugation rule of \cref{eq:z-conjugation} generalizes to
\begin{align}
\label{eq:general-conjugation}
    U_j^\dagger \sigma_v^a U_j
    =
    (-1)^{\eta(a)  g(f(v)) \cdot j}  \sigma_v^a
    ,
    \\
    \eta(a)
    \equiv
    \begin{cases}
        0 & a = x \\
        1 & a \in \{y, z\}
    \end{cases}
    ,
    \nonumber
\end{align}
and the conjugation and averaging steps in the proof of \cref{thm:chadd} apply separately to each Pauli component.

\begin{mycorollary}[Survival rules]
\label{cor:general-survival}
For ideal instantaneous pulses and a time-independent $H_G^\text{gen}$, the cycle-averaged Hamiltonian is
\bes
\begin{align}
\label{eq:general-survival-1local}
    \overline{H}^\text{1-local}
    &=
    \sum_{v \in V} \sum_{a} \delta_{\eta(a)  g(f(v)),  0}  \frac12 h_v^a \sigma_v^a
    ,\\
\label{eq:general-survival-2local}
    \overline{H}^\text{2-local}
    &=
    \sum_{uv \in E} \sum_{a,b} \delta_{\eta(a)  g(f(u)),  \eta(b)  g(f(v))}  \frac14 J_{uv}^{ab} \sigma_u^a \sigma_v^b
    .
\end{align}
\ees
\end{mycorollary}

Thus, to leading order in the Magnus expansion $X$ and $XX$ terms survive every round, for every $g$; a $\sigma^y$ or $\sigma^z$ term survives if and only if its color is assigned to row $i=0$; a two-local term survives if and only if its two factors carry matching row assignments, with an $x$ factor counting as row $0$.
The reason for the invariance of $X$ and $XX$ is that these terms commute with every decoupling operator of the single-axis family and cannot be decoupled or selected independently of the other $X$ and $XX$ terms by it.
A $y$-type single-axis sequence, obtained by replacing $\widetilde{X}_c$ with $\widetilde{Y}_c$ in \cref{eq:chadd-decoupling-operators}, exchanges the roles of $X$ and $Y$ in \cref{eq:general-conjugation}, so the two axes provide different sets of surviving terms. Extracting every coefficient would additionally require a schedule that separates those terms and suitable state preparation and readout; we leave the construction of that extension to future work.

We emphasize that all our exact CHaDD isolation results apply under ideal pulses.
Residual $XY$ interactions can change the Ramsey spectrum, and the applied CHaDD pulses can modify their effect. In separate qubit drive frames, as assumed in \cref{sec:graph-hamiltonian-noise-model}, transverse couplings generally acquire time-dependent coefficients; their cycle averages require integrating those coefficients together with the pulse signs, so \cref{cor:general-survival} does not apply to them unchanged.
The fit uncertainties we report do not bound systematic errors from these omitted dynamics, and we have not separately quantified this source of error in our Ankaa-3 experiments.

\section{Conclusions and Outlook}
\label{sec:conc}

To summarize, we have presented GraphSpec, a whole-chip method for measuring every qubit detuning and qubit-qubit $ZZ$ coupling frequency of a hardware graph of arbitrary size and connectivity using a number of whole-chip Ramsey settings bounded linearly in the distance-2 chromatic number of the graph.
Complementary to the parallel characterization approaches~\cite{evert2024syncopated,Huang2023learning,Gu2024practical}, CHaDD GraphSpec converts a hardware-graph coloring into a complete, deterministic schedule of whole-chip Ramsey measurements that exactly isolates the desired static $Z$-type coefficients under ideal pulses; five settings suffice on Ankaa-3 to determine the detunings, couplings, and corrective drive-frequency shifts. 
The experiment used ten settings to measure each coupling twice, once from the Ramsey signal of each qubit in the coupled pair.
We demonstrated CHaDD GraphSpec on the 84-qubit Rigetti Ankaa-3 QPU and tested the extracted detunings directly, by applying corrective drive frequencies and confirming that the measured detunings are strongly suppressed; the spread of the detuning estimates of the updated qubits decreased by a factor of 4 in a single whole-chip retuning round.

Our results motivate a number of open problems. Isolating arbitrary one- and two-local Pauli terms with CHaDD requires extensions using more than one pulse axis, and a complete non-$Z$ protocol will also require suitable preparation and readout. Measuring changes in detunings and couplings during gate operation is another direction.
Longer-range interactions can be included by adding their edges to $G$ and constructing a distance-2 coloring of the expanded interaction graph.
Shot allocation and total measurement time remain to be optimized at fixed parameter precision. And finally, an experimental test of coupling correction also remains open, since we could not modify the Ankaa-3 couplings in our experiments.

\acknowledgments
This material is based upon work supported by, or in part by, the U.S. Army Research Laboratory and the U.S. Army Research Office under contract/grant number W911NF2310255.
This research was supported by the Office of the Director of National Intelligence (ODNI), Intelligence Advanced Research Projects Activity (IARPA) and the Army Research Office, under the Entangled Logical Qubits program through Cooperative Agreement Number W911NF-23-2-0216.

\section*{Data Availability}
The raw data, fitting routines, and notebooks needed to recreate the plots presented in this paper are available online~\cite{brown2026graphspecdatarepo}.

\appendix

\section{MFA GraphSpec}
\label[appendix]{sec:mfa-graphspec}
In this appendix we describe an alternative protocol that extracts all qubit detuning and qubit-qubit coupling frequencies of an arbitrary hardware graph in $\mathcal{O}(\chi_d)$ whole-chip Ramsey measurements, without dynamical decoupling: the couplings are exposed by spectator state preparation alone.
We (i) derive the two-qubit frequency shifts, (ii) obtain the multi-qubit Ramsey signal, and (iii) establish the measurement bound.

\subsection{Frequency Shifts via the Mean-Field Approximation}
\label[appendix]{sec:mfa}

Consider a simple two-qubit drive-frame Hamiltonian
\begin{align}
\label{eq:2q-hamiltonian}
    \tilde{H} = \frac12 \Delta_1^\text{det} \sigma_1^z + \frac12 \Delta_2^\text{det} \sigma_2^z + \frac14 J_{12} \sigma_1^z \sigma_2^z.
\end{align}
The $1/4$ coefficient in the interaction term is chosen so that the difference between the effective qubit frequency of qubit $1$ when qubit $2$ is in the excited versus ground states is $J_{12}$, and vice versa.

To describe spectator-induced frequency shifts, we first take the partial expectation of the multi-qubit Hamiltonian over the spectator qubits~\cite{tripathi2021suppression}; the resulting mean-field approximation (MFA) does not generally describe the full reduced dynamics, but will be very useful for our purposes.
Specifically, taking the partial expectation over qubit 2 in the state $\ket{\psi}_2$ yields the effective Hamiltonian
\bes
\label{eq:effective-1q-hamiltonian}
\begin{align}
    \bra{\psi}_2 \tilde{H} \ket{\psi}_2 &= \frac12 \Delta_1^\text{det} \sigma_1^z + \frac12 \Delta_2^\text{det} \left< \sigma_2^z \right> + \frac14 J_{12} \sigma_1^z \left< \sigma_2^z \right> \\
\label{eq:effective-1q-detuning-frequency}
    &= \frac{1}{2} \Delta_1^\text{eff} \sigma_1^z ,\quad \Delta_1^\text{eff}\equiv \Delta_1^\text{det} + \frac{1}{2} J_{12} \left< \sigma_2^z \right> ,
\end{align}
\ees
where in the last step we discarded a constant energy shift proportional to the identity.
The effective detuning frequency of qubit 1 is then the precession frequency generated by the resultant Hamiltonian, $\Delta_1^\text{eff}$.

For the eigenstate and equator preparations used below, the carrier frequencies give linear equations relating the detunings to the coupling.
As we show below, this extends to an arbitrary graph of qubits and couplings, provided the spectators of several main qubits can be manipulated independently and simultaneously. If all qubits are prepared in the $\ket{+}$ state, then every qubit has all of its neighbors in a $\langle \sigma^z \rangle = 0$ state, and each qubit's effective detuning is unaffected. If, on the other hand, a set of sufficiently distant qubits is prepared in the $\ket{0}$ state, then the effective detuning of every qubit coupled to one of them is shifted by $+\tfrac12 J_{uv}$ relative to its baseline value, with $J_{uv}$ the corresponding coupling.

\subsection{Multi-Qubit Ramsey Experiment}
\label[appendix]{sec:multi-qubit-ramsey}

Next, we describe a multi-qubit extension of the single-qubit Ramsey experiment (\cref{sec:ramsey}).
To do so, we first develop a multi-qubit version of the MFA in \cref{eq:effective-1q-hamiltonian}.
To describe the Ramsey signal of qubit $v$, define the neighborhood Hamiltonian $\widetilde{H}_v^\mathrm{neigh}$ by retaining the one-qubit terms in $N_v$ [\cref{eq:N_v}] and the couplings incident on $v$:
\begin{align}
\label{eq:neighborhood-hamiltonian}
    \widetilde{H}_v^\mathrm{neigh} &= \frac12 \Delta_v^\text{det} \sigma_v^z + \sum_{\substack{u \in V \\ d(u,v)=1}} \left( \frac12 \Delta_u^\text{det} \sigma_u^z + \frac14 J_{uv} \sigma_u^z \sigma_v^z \right) .
\end{align}
The omitted terms act only on spectator qubits and commute with this Hamiltonian, so they cancel under the partial trace used to obtain the state of qubit $v$.

To determine the dependence of qubit $v$'s frequency on its spectator qubits' states, we take the partial expectation of the above Hamiltonian over the spectator qubits, with spectator $u$ in the state $\ket{\psi_u}$, to arrive at an MFA Hamiltonian for qubit $v$. Let 
\bes
\begin{align}
\ket{\phi_v} &\equiv \bigotimes_{\substack{u \in V \\ d(u,v)=1}} \ket{\psi_u} \\
\label{eq:effective-detuning-frequency}
 \Delta_v^\text{eff} &\equiv \Delta_v^\text{det} + \frac12 \sum_{\substack{u \in V \\ d(u,v)=1}} J_{uv} \left< \sigma_u^z \right> .
\end{align}
\ees
Then,
\beq
    H_v^\text{eff} = \bra{\phi_v} \widetilde{H}_v^\mathrm{neigh} \ket{\phi_v} 
    = \frac12  \Delta_v^\text{eff} \sigma_v^z ,
\eeq
where we discarded a constant term, and we identify $\Delta_v^\text{eff}$ as the effective detuning frequency of qubit $v$ depending on its spectator qubits' states.

Within the MFA, qubit $v$ precesses about the $z$-axis with angular frequency $\Delta_v^\text{eff}$, giving the Ramsey signal of \cref{eq:ramsey-frequency} with $\Delta_v^\text{det}$ replaced by $\Delta_v^\text{eff}$. Note that for spectator qubits prepared in $\sigma^z$ eigenstates the mean-field treatment is exact, since every term of $\tilde{H}_G$ commutes with every $\sigma_u^z$: conditioned on the spectator eigenvalues $\left< \sigma_u^z \right> = \pm1$, qubit $v$ precesses at exactly $\Delta_v^\text{det} + \frac12 \sum_u J_{uv} \left< \sigma_u^z \right>$.

The exact reduced dynamics can also contain spectator-induced modulation, as follows.
Let us assume that every spectator is prepared either in a $\sigma^z$ eigenstate or on the equator, and let 
\beq
N_v^0 \equiv \{ u \in N_v \setminus \{v\} ~|~ \left<\sigma_u^z\right> = 0 \}
\eeq 
denote the set of equator spectators of qubit $v$. Conditioned on their eigenvalues $s_u = \pm1$, which we collect in $s \equiv (s_u)_{u \in N_v^0}$, qubit $v$ precesses at the conditional detuning 
\beq
\Delta_v(s) \equiv \Delta_v^\text{eff} + \frac{1}{2} \sum_{u \in N_v^0} J_{uv} s_u,
\eeq 
so the spectator-averaged closed-system fidelity of the $\ket{+}$ state is
\begin{align}
\label{eq:ramsey-envelope}
    \overline{F}_+
    &\equiv
    \frac{1}{2^{|N_v^0|}}
    \sum_{\{s_u = \pm1\}_{u \in N_v^0}}
    \left\{ \frac12 + \frac12 \cos\left[ \left( \omega_a + \Delta_v(s) \right) t \right] \right\}
    \nonumber \\
    &=
    \frac12 + \frac12 \cos\left[ (\omega_a + \Delta_v^\text{eff}) t \right] \prod_{u \in N_v^0} \cos\left( \tfrac12 J_{uv} t \right)
    .
\end{align}
Thus, an equator spectator leaves the carrier unchanged but produces beating and revivals through its cosine envelope factor.

\subsection{Reconstruction and Measurement Bound}
\label[appendix]{sec:mfa-reconstruction}

\begin{mytheorem}[MFA GraphSpec]
\label{thm:graph-spectroscopy}
$\chi_d+1$ whole-chip Ramsey measurements, with an artificial detuning satisfying
\beq
\label{eq:omega-a-bound}
\omega_a > \max_{v \in V} \Big( |\Delta_v^\text{det}| + \frac12 \sum_{u:uv \in E} |J_{uv}| \Big)
\eeq
and resolved sideband frequencies, are sufficient to measure every qubit detuning frequency $\Delta_v^\text{det}$ and every qubit-qubit $ZZ$ coupling frequency $J_{uv}$ on a hardware graph $G=(V,E)$ with distance-$d$ chromatic number $\chi_d$ for $d \geq 2$.
\end{mytheorem}

\begin{proof}
Perform a baseline whole-chip Ramsey experiment with artificial detuning $\omega_a$ in which all qubits are prepared in the $\ket{+}$ state regardless of color. Each qubit $v$ then has $\left<\sigma_u^z\right>=0$ for all of its neighbors $u$, and so has effective detuning $\Delta_v^\text{eff}=\Delta_v^\text{det}$ by \cref{eq:effective-detuning-frequency}. The Ramsey signal on each qubit $v \in V$ then has carrier frequency $\omega_a + \Delta_v^\text{det}$, which the envelope factors of \cref{eq:ramsey-envelope} split into sidebands at $\omega_a + \Delta_v^\text{det} + \frac12 \sum_u s_u J_{uv}$ with $s_u = \pm 1$; under \cref{eq:omega-a-bound} every sideband frequency is positive and the carrier is the midpoint of the lowest and highest sidebands, which determines every detuning $\Delta_v^\text{det}$.

Now, pick a color $c$ in the distance-$d$ coloring, and set all qubits of that color to be in the $\ket{0}$ state and every other qubit of color $c' \neq c$ in the $\ket{+}$ state.
Recall that since $d \geq 2$, all qubits of color $c' \neq c$ have at most one neighbor $u$ of color $c$ (\cref{lem:neighborhood-distinct-colors}), i.e., with $\left<\sigma_u^z\right>=1$.
Then for $v \in V$ with $f(v) = c' \neq c$,
\begin{align}
\label{eq:one-color-frequency-shift}
    \Delta_v^\text{eff} = \begin{cases}
        \Delta_v^\text{det} + \frac12 J_{uv} & \text{$v$ has a neighbor $u$ with } f(u)=c \\
        \Delta_v^\text{det} & \text{otherwise}
    \end{cases}
    ,
\end{align}
according to \cref{eq:effective-detuning-frequency}.
A whole-chip Ramsey experiment with artificial detuning $\omega_a$ and the qubits prepared in this fashion then yields, on each qubit $v$ of color $c' \neq c$, Ramsey signals with carrier frequency
$\omega_a + \Delta_v^\text{eff}$, with $\Delta_v^\text{eff}$ as in \cref{eq:one-color-frequency-shift}; since the selected color is prepared in $\sigma^z$ eigenstates, its contribution to the carrier is exact rather than a mean-field approximation (as argued in \cref{sec:multi-qubit-ramsey}), while the remaining neighbors contribute the envelope factors of \cref{eq:ramsey-envelope}.
Repeat this for each of the $\chi_d$ colors $c = 0, \dots, \chi_d - 1$.

The resulting equations are solved in two steps.
The baseline experiment determines $\Delta_v^\text{det}$ for every $v \in V$.
For any edge $uv \in E$, the experiment in which color $f(v)$ is prepared in $\ket{0}$ has carrier frequency $y_u^{(f(v))} = \omega_a + \Delta_u^\text{det} + \frac12 J_{uv}$ on qubit $u$, where $y_u^{(c)}$ denotes the carrier frequency of qubit $u$, extracted as the sideband midpoint, in the experiment in which color $c$ is prepared in $\ket{0}$; hence
\begin{align}
\label{eq:mfa-coupling-extraction}
    J_{uv}
    =
    2 \left( y_u^{(f(v))} - \omega_a - \Delta_u^\text{det} \right)
    .
\end{align}
Every coupling is thus determined explicitly.
In fact, each $J_{uv}$ is determined twice, once from each endpoint (from $y_u^{(f(v))}$ and from $y_v^{(f(u))}$), which provides a consistency check and can be used for averaging.
In total, the $\chi_d+1$ experiments determine all $|V|+|E|$ frequencies.
\end{proof}

The number of whole-chip Ramsey experiments in \cref{thm:graph-spectroscopy} is an upper bound: every edge has endpoints of two different colors, so omitting any one selected color leaves each edge measured at an endpoint of another color, and the baseline plus $\chi_d - 1$ selected-color experiments already determine every parameter.
The requirement $d \geq 2$ enters through \cref{lem:neighborhood-distinct-colors}: it guarantees that each qubit has at most one spectator-color neighbor, so its frequency shift isolates a single coupling.

Proof-of-concept experiments with MFA GraphSpec appear in \cref{sec:mfa-graphspec-experiments}, and a comparison with CHaDD GraphSpec for detuning correction appears in \cref{app:comparison}.
MFA GraphSpec uses no pulses during the idle period and therefore avoids errors from the added CHaDD pulses, though preparation and readout errors remain. Its undecoupled spectator couplings modulate the Ramsey contrast through the cosine factors in \cref{eq:ramsey-envelope}, complicating carrier-frequency estimation.
For Ankaa-3, a single ideal baseline experiment suffices for the detunings alone [the first step of the proof of \cref{thm:graph-spectroscopy}], versus $\chi = 2$ whole-chip experiments for CHaDD GraphSpec; the full set of detunings and couplings takes $\chi_2 = 5$ experiments (see the remark after the proof of \cref{thm:graph-spectroscopy}) for both methods.
Choosing between the protocols requires balancing errors from the added pulses against uncertainty and bias in estimating carriers with spectator-induced modulation.

\section{MFA GraphSpec Experiments}
\label[appendix]{sec:mfa-graphspec-experiments}

\begin{figure*}
\centering
\includegraphics[width=0.98\textwidth]{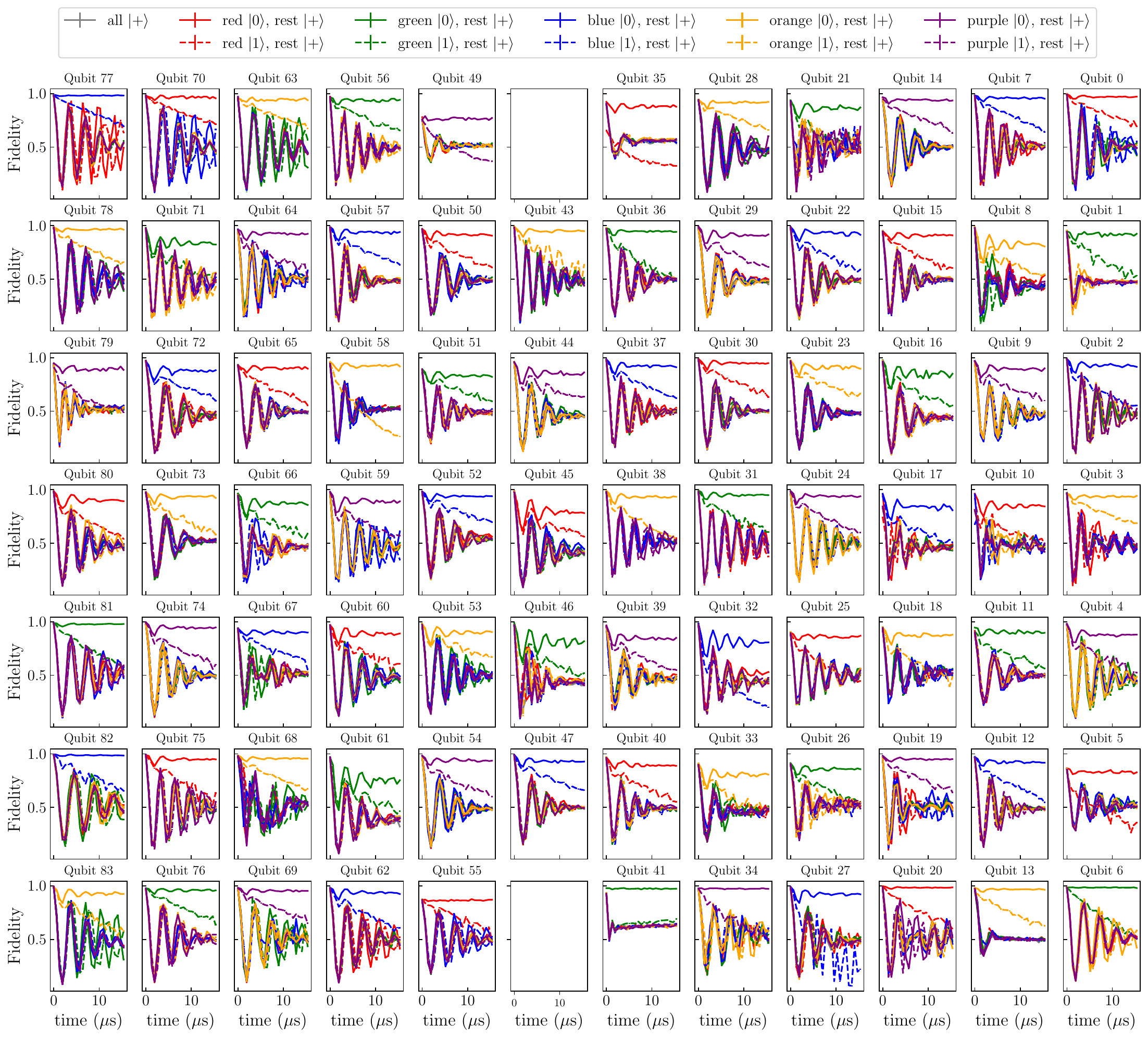}
\caption{
All qubits' data for the 11 whole-chip MFA GraphSpec experiments.
The qubits' panels are laid out as the qubits are in \cref{fig:ankaa3-distance2-5coloring}.
Blank panels correspond to the inoperable qubits 42 and 48.
Gray lines correspond to the Ramsey results from the baseline experiment in which all qubits are prepared in the $\ket{+}$ state.
Solid (dashed) red lines show the results when red qubits are prepared in $\ket{0}$ ($\ket{1}$) and all other qubits are prepared in $\ket{+}$;
for a non-red qubit $v$ with a red neighbor $u$, these preparations shift the effective detuning by $+J_{uv}/2$ ($-J_{uv}/2$).
For a red qubit, such as qubit 80, the red curves instead record populations because that qubit is prepared in a $\sigma^z$ eigenstate rather than a Ramsey superposition.
The same distinction applies to the other colors: in each qubit's panel, the two curves matching its assigned color in \cref{fig:ankaa3-distance2-5coloring} are population records, while the remaining curves are Ramsey signals.
}
\label{fig:ankaa3-graph-spec-all-qubits}
\end{figure*}

\begin{figure}
\centering
\includegraphics[width=0.98\columnwidth]{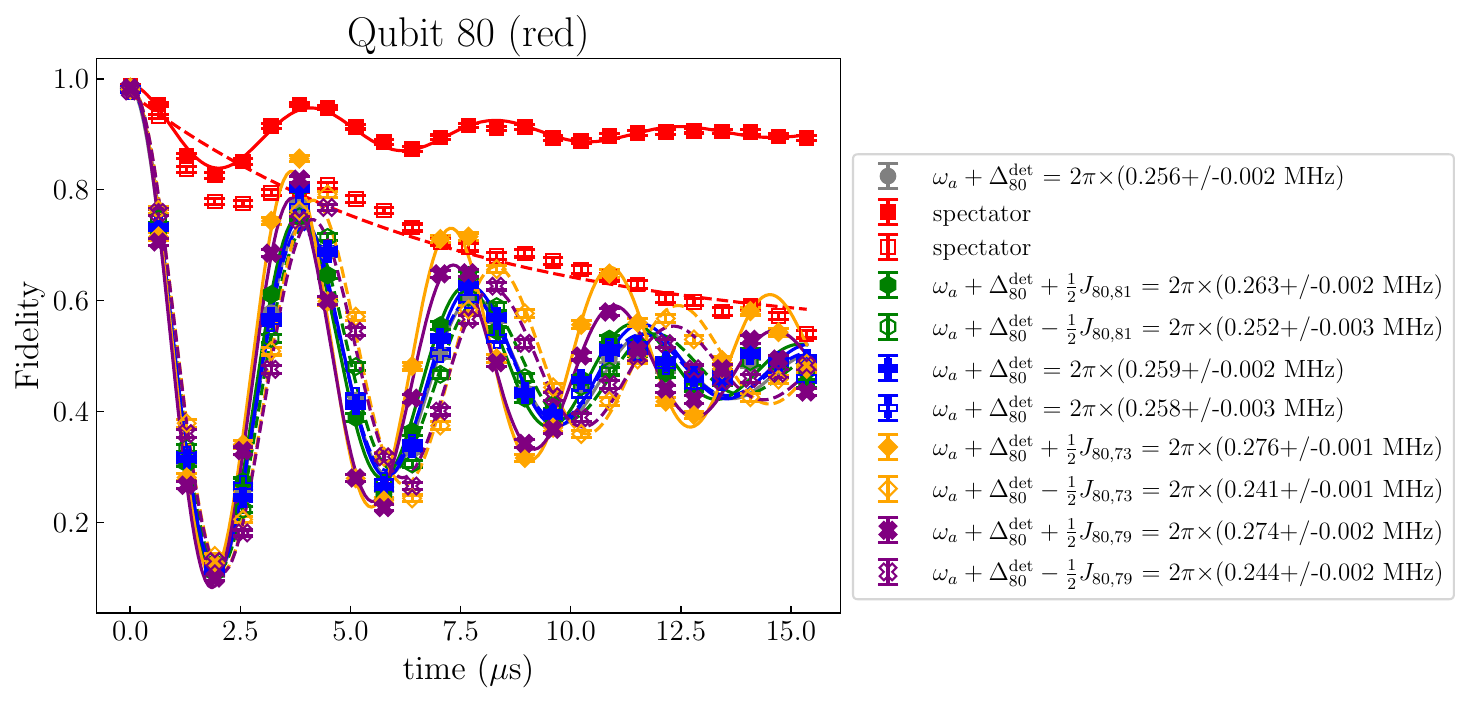}
\caption{
Qubit 80's data from the 11 whole-chip MFA GraphSpec experiments. As shown in \cref{fig:ankaa3-distance2-5coloring}(b), qubit 80 is a red qubit with purple, orange, and green neighbors to the north, east, and south; unlike red qubits in the bulk, it lacks a blue neighbor to the west. 
In the $Z$-type model, preparing the blue qubits in $\ket{0}$ or $\ket{1}$ leaves qubit 80's Ramsey signal unchanged; the corresponding fitted frequencies are consistent with the gray baseline within the reported uncertainties, and we leave them out of the linear system of \cref{eq:wls-system}, since for this qubit they give the same equation as the baseline experiment, $\Delta_v^\text{eff}=\Delta_v^\text{det}$ [\cref{eq:effective-detuning-frequency}].
The solid-dashed fitted-frequency differences are $35$, $30$, and $11$~kHz for the orange, purple, and green neighbors, respectively.
}
\label{fig:ankaa3-graph-spec-qubit-80}
\end{figure}

We performed 11 whole-chip MFA GraphSpec experiments, with preparation-state settings determined by the distance-2 5-coloring of Ankaa-3 depicted in \cref{fig:ankaa3-distance2-5coloring}(b).
A set of $\chi_2+1=6$ experiments suffices to extract all detunings and couplings (\cref{thm:graph-spectroscopy,cor:square-lattice-chromatic-numbers}): the baseline and the five experiments, depicted with solid lines in \cref{fig:ankaa3-graph-spec-all-qubits}, in which the spectator qubits occupy the $\ket{0}$ state.
We also performed an additional $\chi_2=5$ experiments in which the spectator qubits occupy the $\ket{1}$ state instead, depicted with dashed lines in \cref{fig:ankaa3-graph-spec-all-qubits}.
The additional experiments are largely redundant, but they can reveal deviations from the assumed spectator values $\langle\sigma^z\rangle = \pm1$, e.g., due to state-preparation errors.
Throughout \cref{fig:ankaa3-graph-spec-all-qubits,fig:ankaa3-graph-spec-qubit-80}, the color of a plotted line indicates the color of the qubits treated as spectators, and a solid (dashed) line indicates spectators prepared in $\ket{0}$ ($\ket{1}$); the gray lines are the baseline.
\Cref{fig:ankaa3-graph-spec-qubit-80} focuses on qubit 80 (red); the frequency difference between the solid and dashed curves of a given color corresponds to the coupling between this qubit and its neighbor of that color, which is substantial here for qubits 73 (orange) and 79 (purple).

The baseline experiment, in which all qubits are prepared in the $\ket{+}$ state, is plotted in gray; it measures the detuning of every qubit simultaneously, while the couplings do not shift the carrier frequencies; their magnitudes enter only through the envelope factors of \cref{eq:ramsey-envelope}.
When all red qubits are prepared in the $\ket{0}$ state and the remainder in $\ket{+}$, the resulting Ramsey fringes are plotted in solid red.
Relative to the gray baseline, the effective detuning of each non-red qubit with a red neighbor is shifted up by half the qubit-qubit coupling to its sole red neighbor [\cref{eq:one-color-frequency-shift}]; comparing these two experiments therefore determines roughly $2/5$ of the qubit-qubit couplings on the chip.
The same couplings are determined by the complementary experiment, plotted as dashed red, in which all red qubits are prepared in the $\ket{1}$ state instead: in that case, the effective detuning of every non-red qubit with a red neighbor is shifted down by the same amount.
Incidentally, these two experiments also perform a $T_1$-like measurement on the red qubits themselves, in the presence of crosstalk, since each red qubit idles in $\ket{0}$ or $\ket{1}$ with all of its neighbors on the equator.

Within the $Z$-type model, a larger frequency difference between the solid and dashed curves of a given color indicates a larger coupling magnitude to the neighbor of that color in \cref{fig:ankaa3-graph-spec-all-qubits}.
For example, qubits 66 (green) and 67 (blue) share a substantial $ZZ$ coupling, as evidenced by the difference between the solid and dashed blue curves in qubit 66's panel and between the solid and dashed green curves in qubit 67's panel.
An individual qubit's panel likewise identifies the neighbor with which it shares the largest coupling: qubit 80's orange and purple neighbors, qubits 73 and 79, couple to it with comparable strength, as indicated by the comparable solid-dashed splittings of those two colors in \cref{fig:ankaa3-graph-spec-qubit-80}.

Having collected data that overdetermine the qubit detuning and qubit-qubit coupling frequencies on the chip, we turn to extracting them.
For each of the 82 qubits, we fit the nine curves in which it is prepared in $\ket{+}$ to a decaying cosine and extract the frequency and its uncertainty.
The other two curves are spectator-population records and enter no frequency equation.
Note that the envelope factors of \cref{eq:ramsey-envelope} can displace this fitted frequency from the carrier, which is a systematic effect the fit uncertainty does not capture.
We then form the overdetermined linear system
\begin{align}
\label{eq:wls-system}
    A x = b
    ,
\end{align}
where $x \in \mathbb{R}^{|V|+|E|}$ collects the unknown detunings $\Delta_v^\text{det}$ and couplings $J_{uv}$, and $b$ collects the corresponding fitted Ramsey frequencies minus $\omega_a$, with uncertainties $\delta b$.
Each row of $A$ corresponds to one retained (qubit, experiment) pair in which the qubit is prepared in $\ket{+}$ and encodes \cref{eq:effective-detuning-frequency} for that pair, i.e., it has a $1$ in the column of $\Delta_v^\text{det}$ and, for the experiments other than the baseline (non-gray in our plots), $\frac12 \left<\sigma_u^z\right> \in \{0, \pm\frac12\}$ in the column of $J_{uv}$ for each neighbor $u$ of $v$, as determined by the preparation-state settings of that experiment.
For a qubit on the boundary that lacks a neighbor of a given color, e.g., qubit 80 (red), which lacks a blue neighbor [\cref{fig:ankaa3-graph-spec-qubit-80}], the two experiments of that color give the same equation as the baseline experiment, $\Delta_v^\text{eff}=\Delta_v^\text{det}$ [\cref{eq:effective-detuning-frequency} with $\left<\sigma_u^z\right>=0$ for every neighbor $u$], and are not included in the system.
The system has more equations than unknowns, and the fitted frequencies carry noise, so we solve it in the weighted least squares sense, which both propagates the fit uncertainties to the extracted frequencies and can reduce the propagated uncertainties relative to unweighted least squares.
Since fits to experimental data occasionally fail, due to a poorly performing qubit or inadequate initial fit parameters, the overdetermination provides a useful redundancy when individual curves fail to fit. If all probe curves of a qubit fail, its detuning is unrecoverable; recovery of the remaining parameters requires removing that unknown and checking the rank of the retained system.

Assuming that $A$ has full column rank after any exclusions, and that the fitted-frequency errors are unbiased with covariance $\mathrm{diag}(\delta b_i^2)$, we construct the diagonal weight matrix $W = \mathrm{diag}(1/\delta b_i^2)$ from the inverse variances of the fitted Ramsey frequencies, and compute the covariance matrix of the solution, $\Sigma_x = \left(A^T W A\right)^{-1}$, the weighted least squares solution $x^* = \Sigma_x A^T W b$, and the propagated one-standard-deviation errors $\sigma_{x_i} = \sqrt{(\Sigma_x)_{ii}}$.

The system has $|V|+|E|=225$ unknowns: the detunings of the $|V|=82$ operable qubits and the $|E|=143$ couplings between them, including those on the five edges whose two-qubit gates are unavailable.
To keep failed fits from skewing the solution, we admitted a fitted frequency $\hat\omega$ into the system only if the fit returned an uncertainty for it and $|\hat\omega-\omega_a|\leq0.33\,\omega_a$, i.e., only if it lay within $99$~kHz of the artificial detuning.
No fit of qubit 41 or of qubit 79 was admitted, and we removed these two qubits from the system.
They therefore received no update in the detuning-correction test of \cref{app:comparison} below and are omitted from both histograms in \cref{fig:mfa-detuning-correction}; qubit 46, none of whose post-correction fits was admitted, is omitted from \cref{fig:mfa-detuning-correction}(b).

\section{Comparison of CHaDD GraphSpec and MFA GraphSpec for Detuning Correction}
\label[appendix]{app:comparison}

\begin{figure}
\centering
(a)
\includegraphics[width=0.98\columnwidth]{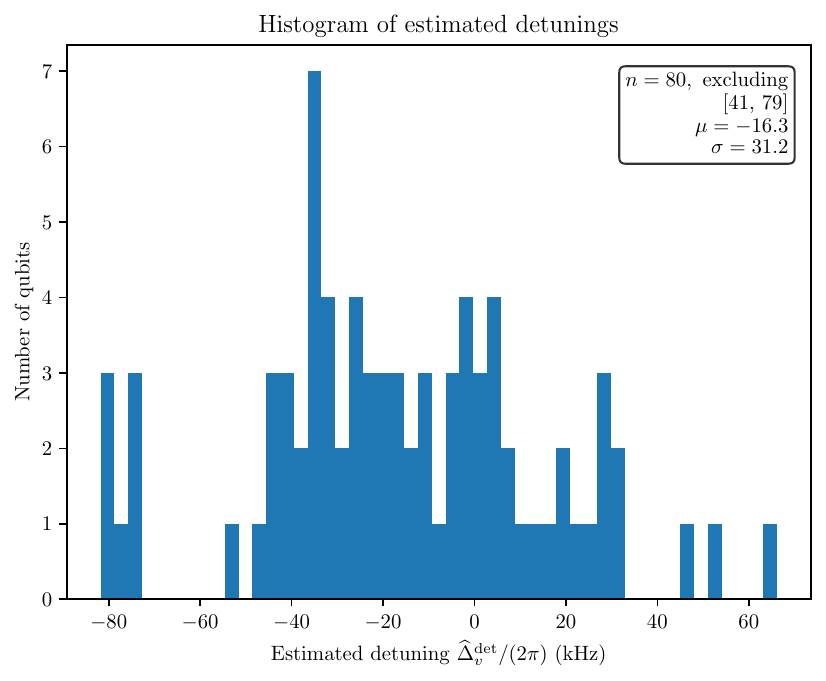}
\\
(b)
\includegraphics[width=0.98\columnwidth]{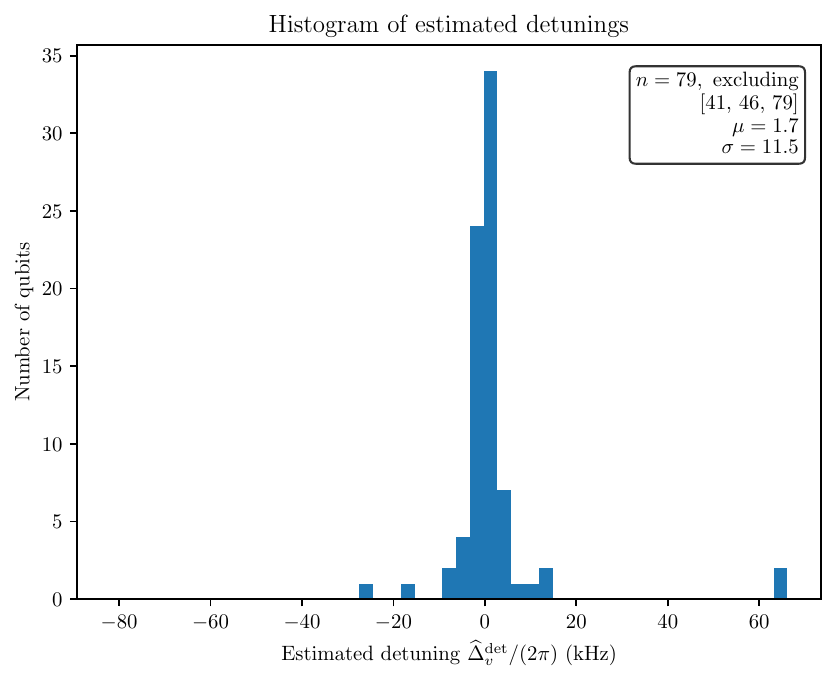}
\caption{
Detuning correction with MFA GraphSpec, repeating the experiment of \cref{fig:chadd-detuning-correction}.
Histograms of the measured qubit detunings across the chip for (a) the diagnostic scan and (b) the post-correction scan.
Note the different horizontal scale and bin width from \cref{fig:chadd-detuning-correction}; the annotations are as in that figure.
}
\label{fig:mfa-detuning-correction}
\end{figure}

To compare the two GraphSpec variants, we repeated the detuning-correction experiment of \cref{sec:detuning-correction} with MFA GraphSpec; the corresponding diagnostic and post-correction histograms appear in \cref{fig:mfa-detuning-correction}.
We ran the four scans of \cref{fig:chadd-detuning-correction,fig:mfa-detuning-correction} within an hour of one another, so as to minimize the effect of drift, and applied each protocol's updates to the original drive frequencies, so that the two tests are independent.
As with CHaDD GraphSpec (\cref{sec:detuning-correction}), the MFA GraphSpec updates applied at the time of the experiment were computed with an earlier version of the fitting routine;
for each of the 80 qubits that received an update, the applied update agrees with the diagnostic estimate shown in \cref{fig:mfa-detuning-correction}(a) to within 2~kHz, except for qubit 66, for which the two agree to within 6~kHz.
MFA GraphSpec reduces the standard deviation of the measured detunings by a factor of 2.7 (from 31.2 to 11.5~kHz) and moves the mean from $-16.3$ to $1.7$~kHz;
by comparison, CHaDD GraphSpec reduces the standard deviation by a factor of 3.8 (from 127.4 to 33.7~kHz) and moves the mean from $-45.6$ to $3.8$~kHz.
In both cases a few qubits dominate the post-correction standard deviation: qubit 39 for CHaDD GraphSpec (\cref{sec:detuning-correction}), and qubits 49 and 56, both near $65$~kHz, for MFA GraphSpec; without them it is 8.7 and 5.0~kHz, respectively.

The two tests also differ in which detunings they can register.
The MFA GraphSpec scans read out the population of $\ket{+}$, which does not distinguish the two signs of $\omega_a+\Delta_v^\text{det}$ (\cref{sec:ramsey}), and fitted frequencies further than $99$~kHz from $\omega_a$ are not admitted into the linear system (\cref{sec:mfa-graphspec-experiments}), so that a detuning of larger magnitude cannot be registered.
The CHaDD GraphSpec scans of \cref{fig:chadd-detuning-correction}, which read out the population of $\ket{+y}$, are subject to neither restriction.
Two qubits illustrate the difference.
The fringes of qubit 79 are well resolved, with fitted frequencies about $130$~kHz above $\omega_a$, i.e., outside this window; it has no MFA GraphSpec estimate, whereas CHaDD GraphSpec estimates its detuning at $120$~kHz in the diagnostic scan and at $-2.6$~kHz in the post-correction scan.
For qubit 46 the three best-determined fitted frequencies of the MFA GraphSpec scan lie between $0.41$ and $0.43$~MHz, likewise outside the window; they agree with the CHaDD GraphSpec estimate of $-723$~kHz, for which $|\omega_a+\Delta_v^\text{det}|=2\pi\times0.42$~MHz, whereas the MFA GraphSpec estimate of $2.6$~kHz rests on the remaining, less well determined fits.
The two protocols also disagree on qubit 49, to which MFA GraphSpec assigns $-31$~kHz and CHaDD GraphSpec $-792$~kHz.
The CHaDD GraphSpec estimate is confirmed by the post-correction scan, which finds this qubit at $-1.5$~kHz after its update (\cref{sec:detuning-correction}); at a detuning of $-792$~kHz, however, the MFA GraphSpec diagnostic scan would have found fitted frequencies near $0.49$~MHz, not the $0.24$ to $0.29$~MHz it found.
A possible explanation is that the frequency of this qubit changed between the two diagnostic scans, consistent with the fast decay of its fringes.

\section{Effects of Inoperable Qubits and Missing Edges}
\label[appendix]{app:inoperable-qubits}

In this appendix we distinguish unavailable two-qubit gates from omitted physical interactions, derive the contributions of unpulsed inoperable qubits to the Ramsey signals, and determine their effects on detuning correction and coupling extraction.

\subsection{Unavailable gates and omitted interactions}

An unavailable native two-qubit gate does not imply that the corresponding residual coupling vanishes. GraphSpec requires single-qubit preparation, control, and readout at the endpoints of an interaction, so two-qubit gate unavailability alone does not prevent the edge's measurement. Thus, every physical coupling included in the model between operable qubits belongs in the interaction edge set $E$, irrespective of whether the corresponding two-qubit gate is available. This is how we treat the five unavailable gate edges listed in \cref{sec:experiments}. A genuinely absent interaction, with $J_{uv}=0$, can be omitted without affecting the protocol.

Omitting a nonzero interaction from the graph used for coloring can invalidate the isolation of individual coefficients. For an omitted edge $uv$ between operable qubits, the same conjugation calculation as in \cref{thm:chadd} gives
\begin{equation}
\label{eq:omitted-edge-average}
\overline{H}_{uv}^{\mathrm{int}}
=
\delta_{g(f(u)),g(f(v))}
\frac{J_{uv}}4 \sigma_u^z\sigma_v^z .
\end{equation}
If its endpoints have the same color, this interaction survives every color-to-row map. If they have distinct colors, it survives whenever their colors share a row and can introduce an additional neighbor into a purportedly isolated pair.

For example, suppose the modeled graph contains the edge $va$ and an isolated vertex $b$, while the physical Hamiltonian also contains the omitted edge $vb$. The assignment $f(v)=0$ and $f(a)=f(b)=1$ is a proper distance-2 coloring of the modeled graph. Assigning both colors to the same nonzero row and preparing $v$ in $\ket{+}$ and $a,b$ in $\ket{0}$ yields the Ramsey frequency
\begin{equation}
\label{eq:omitted-edge-example}
\omega_a+\frac12(J_{va}+J_{vb}) .
\end{equation}
Interpreting this frequency as an isolated measurement of $J_{va}$ returns $J_{va}+J_{vb}$. Including the physical edge $vb$ in the coloring graph prevents this assignment, since $a$ and $b$ must then receive different colors.

\subsection{Residual couplings to inoperable qubits}

Let $V_{\mathrm{inop}}$ denote the inoperable qubits omitted from $G=(V,E)$, where $V$ contains the operable qubits. We consider the case in which no control pulses are applied to the omitted qubits. We assume ideal instantaneous pulses on $V$ and the static $Z$-type Hamiltonian
\begin{subequations}
\label{eq:inoperable-full-hamiltonian}
\begin{align}
\widetilde{H}_{\mathrm{full}}
&=
\widetilde{H}_G+H_{\mathrm{inop}}
+\frac12\sum_{v\in V}\sigma_v^z B_v ,
\\
B_v
&\equiv
\frac12\sum_{u\in V_{\mathrm{inop}}}
J_{uv}\sigma_u^z .
\end{align}
\end{subequations}
Here $H_{\mathrm{inop}}$ contains the terms supported entirely on $V_{\mathrm{inop}}$, and we set $J_{uv}=0$ when no interaction is present. We assume that $E$ includes all modeled interactions between operable qubits.

Since the omitted qubits are unpulsed, each boundary interaction acquires only the toggling sign of its operable endpoint. Averaging over a completed CHaDD cycle therefore gives
\begin{equation}
\label{eq:inoperable-average}
\overline{H}_{\mathrm{full}}
=
\overline{H}_G+H_{\mathrm{inop}}
+\frac12\sum_{v\in V}
\delta_{g(f(v)),0}\sigma_v^z B_v .
\end{equation}
All toggled Hamiltonians commute, so this average generates the exact evolution at completed cycles. Boundary interactions survive on the row-0 qubits used to measure detunings and vanish on qubits assigned to nonzero rows.

Consider a row-0 qubit $v$ prepared in $\ket{+}$, initially uncorrelated with the remaining qubits, and let $\rho_{\mathrm{inop}}$ be the arbitrary initial state of the omitted qubits. Once interactions between $v$ and the other operable qubits have been decoupled, its exact Ramsey signal at completed cycles is
\begin{equation}
\label{eq:inoperable-ramsey}
F_{+,v}(t)
=
\frac12+\frac12
\operatorname{Re}\left[
e^{-i(\omega_a+\Delta_v^{\mathrm{det}})t}
\operatorname{Tr}\left(
\rho_{\mathrm{inop}}e^{-itB_v}
\right)
\right].
\end{equation}

If the omitted spectators occupy fixed $\sigma_u^z$ eigenstates with eigenvalues $s_u=\pm1$, define
\begin{equation}
\label{eq:inoperable-detuning-shift}
\delta_v^{\mathrm{inop}}(s)
\equiv
\frac12\sum_{u\in V_{\mathrm{inop}}}J_{uv}s_u .
\end{equation}
The measured detuning is then
$\Delta_v^{\mathrm{det}}+\delta_v^{\mathrm{inop}}(s)$.
For one omitted spectator in $\ket{0}$, this shift is $J_{uv}/2$. If the same spectator is instead prepared on the equator, \cref{eq:inoperable-ramsey} becomes
\begin{equation}
\label{eq:inoperable-equator-signal}
F_{+,v}(t)
=
\frac12+\frac12
\cos[(\omega_a+\Delta_v^{\mathrm{det}})t]
\cos(J_{uv}t/2).
\end{equation}
Thus, an omitted spectator can produce either a conditional frequency shift or additional modulation, depending on its state.

\subsection{Consequences for calibration and coupling extraction}

We consider frequency retuning and coupling extraction under the static $Z$-type model of \cref{eq:inoperable-full-hamiltonian}, with ideal instantaneous pulses on the operable qubits.

\begin{enumerate}
\item[(i)] \emph{Frequency retuning can succeed without determining the isolated-qubit frequency}:
For fixed omitted-spectator eigenvalues $s_u=\pm1$, the measured detuning contains both the isolated-qubit detuning and the spectator-induced shift:
$\Delta_v^{\mathrm{det}}+\delta_v^{\mathrm{inop}}(s)$.
Applying the correction
\begin{equation}
\omega_v^{d,\mathrm{new}}
=
\omega_v^{d,\mathrm{old}}
-
[\Delta_v^{\mathrm{det}}+\delta_v^{\mathrm{inop}}(s)]
\end{equation}
makes a repeat measurement with the same spectator states return zero, although the detuning relative to the isolated-qubit frequency is now $-\delta_v^{\mathrm{inop}}(s)$.
The correction therefore calibrates the transition for that spectator configuration.

This distinction matters when the omitted spectators change state.
If one spectator flips from $s_u$ to $-s_u$, the corrected transition acquires a detuning of $-J_{uv}s_u$.
Thus, the experiment in \cref{sec:detuning-correction} is a self-consistency test: an unchanged spectator shift is compensated by the correction and is not distinguished from an isolated-qubit detuning by the repeated measurement.

\item[(ii)] \emph{CHaDD coupling measurements do not require knowledge or preparation of the omitted-qubit states}:
When measuring a coupling between two operable qubits, we assign their colors to a common nonzero Hadamard row.
By \cref{eq:inoperable-average}, this choice retains the desired interaction while canceling the couplings from both endpoints to every unpulsed omitted qubit.
The cancellation is exact at completed cycles and independent of the omitted-qubit state.

Consequently, an inoperable neighbor can affect the detuning correction in item~(i) without contaminating the CHaDD estimate of a coupling between operable qubits.
Under the stated assumptions, that coupling measurement does not require control or characterization of the inoperable neighbor.

\item[(iii)] \emph{MFA coupling extraction can cancel an unchanged spectator contribution, but single-cosine fits do not guarantee this cancellation}:
In MFA GraphSpec, couplings to omitted qubits remain in the Ramsey signals.
Their contribution can instead cancel when we subtract suitable frequencies from two settings, provided the omitted-qubit state is unchanged and the relevant sidebands are positive and resolved.

Let $m_v^{\mathrm{base}}$ denote the midpoint of the lowest and highest sideband frequencies in the baseline setting, and let $m_v^{(w,0)}$ denote the corresponding midpoint when the selected operable neighbor $w$ is prepared in $\ket{0}$.
The unchanged omitted-qubit state contributes the same shift to both midpoints, while preparing $w$ shifts the midpoint by $J_{wv}/2$.
Thus,
\begin{equation}
J_{wv}
=
2\left[m_v^{(w,0)}-m_v^{\mathrm{base}}\right],
\end{equation}
even when the omitted-qubit state is mixed.

The experimental qualification is that a single-cosine fit need not return either midpoint.
Changes in the Ramsey modulation between settings can therefore bias the difference of fitted frequencies, even when the omitted-qubit state is unchanged.
A change in that state can also change the midpoint itself.
The cancellation result consequently supports the resolved-sideband reconstruction, but not an unrestricted subtraction of single-cosine fit frequencies.
\end{enumerate}

These conclusions rely on conservation of the omitted spectators' $Z$ values during each Ramsey evolution.
Relaxation or non-$Z$ dynamics can invalidate the exact CHaDD cancellation in item~(ii) and require an extension of the model that is beyond the scope of this work.

\bibliographystyle{apsrev4-2}
\bibliography{refs}

\begin{thebibliography}{24}\makeatletter
\providecommand \@ifxundefined [1]{\@ifx{#1\undefined}
}\providecommand \@ifnum [1]{\ifnum #1\expandafter \@firstoftwo
 \else \expandafter \@secondoftwo
 \fi
}\providecommand \@ifx [1]{\ifx #1\expandafter \@firstoftwo
 \else \expandafter \@secondoftwo
 \fi
}\providecommand \natexlab [1]{#1}\providecommand \enquote  [1]{``#1''}\providecommand \bibnamefont  [1]{#1}\providecommand \bibfnamefont [1]{#1}\providecommand \citenamefont [1]{#1}\providecommand \href@noop [0]{\@secondoftwo}\providecommand \href [0]{\begingroup \@sanitize@url \@href}\providecommand \@href[1]{\@@startlink{#1}\@@href}\providecommand \@@href[1]{\endgroup#1\@@endlink}\providecommand \@sanitize@url [0]{\catcode `\\12\catcode `\$12\catcode
  `\&12\catcode `\#12\catcode `\^12\catcode `\_12\catcode `\%12\relax}\providecommand \@@startlink[1]{}\providecommand \@@endlink[0]{}\providecommand \url  [0]{\begingroup\@sanitize@url \@url }\providecommand \@url [1]{\endgroup\@href {#1}{\urlprefix }}\providecommand \urlprefix  [0]{URL }\providecommand \Eprint [0]{\href }\providecommand \doibase [0]{https://doi.org/}\providecommand \selectlanguage [0]{\@gobble}\providecommand \bibinfo  [0]{\@secondoftwo}\providecommand \bibfield  [0]{\@secondoftwo}\providecommand \translation [1]{[#1]}\providecommand \BibitemOpen [0]{}\providecommand \bibitemStop [0]{}\providecommand \bibitemNoStop [0]{.\EOS\space}\providecommand \EOS [0]{\spacefactor3000\relax}\providecommand \BibitemShut  [1]{\csname bibitem#1\endcsname}\let\auto@bib@innerbib\@empty
\bibitem [{\citenamefont {Burnett}\ \emph {et~al.}(2019)\citenamefont
  {Burnett}, \citenamefont {Bengtsson}, \citenamefont {Scigliuzzo},
  \citenamefont {Niepce}, \citenamefont {Kudra}, \citenamefont {Delsing},\ and\
  \citenamefont {Bylander}}]{Burnett2019decoherence}\BibitemOpen
  \bibfield  {author} {\bibinfo {author} {\bibfnamefont {J.~J.}\ \bibnamefont
  {Burnett}}, \bibinfo {author} {\bibfnamefont {A.}~\bibnamefont {Bengtsson}},
  \bibinfo {author} {\bibfnamefont {M.}~\bibnamefont {Scigliuzzo}}, \bibinfo
  {author} {\bibfnamefont {D.}~\bibnamefont {Niepce}}, \bibinfo {author}
  {\bibfnamefont {M.}~\bibnamefont {Kudra}}, \bibinfo {author} {\bibfnamefont
  {P.}~\bibnamefont {Delsing}},\ and\ \bibinfo {author} {\bibfnamefont
  {J.}~\bibnamefont {Bylander}},\ }\href
  {https://doi.org/10.1038/s41534-019-0168-5} {\bibfield  {journal} {\bibinfo
  {journal} {npj Quantum Information}\ }\textbf {\bibinfo {volume} {5}},\
  \bibinfo {pages} {54} (\bibinfo {year} {2019})}\BibitemShut {NoStop}\bibitem [{\citenamefont {Klimov}\ \emph {et~al.}(2018)\citenamefont {Klimov},
  \citenamefont {Kelly}, \citenamefont {Chen}, \citenamefont {Neeley},
  \citenamefont {Megrant}, \citenamefont {Burkett}, \citenamefont {Barends},
  \citenamefont {Arya}, \citenamefont {Chiaro}, \citenamefont {Chen},
  \citenamefont {Dunsworth}, \citenamefont {Fowler}, \citenamefont {Foxen},
  \citenamefont {Gidney}, \citenamefont {Giustina}, \citenamefont {Graff},
  \citenamefont {Huang}, \citenamefont {Jeffrey}, \citenamefont {Lucero},
  \citenamefont {Mutus}, \citenamefont {Naaman}, \citenamefont {Neill},
  \citenamefont {Quintana}, \citenamefont {Roushan}, \citenamefont {Sank},
  \citenamefont {Vainsencher}, \citenamefont {Wenner}, \citenamefont {White},
  \citenamefont {Boixo}, \citenamefont {Babbush}, \citenamefont {Smelyanskiy},
  \citenamefont {Neven},\ and\ \citenamefont {Martinis}}]{Klimov:2018ww}\BibitemOpen
  \bibfield  {author} {\bibinfo {author} {\bibfnamefont {P.~V.}\ \bibnamefont
  {Klimov}}, \bibinfo {author} {\bibfnamefont {J.}~\bibnamefont {Kelly}},
  \bibinfo {author} {\bibfnamefont {Z.}~\bibnamefont {Chen}}, \bibinfo {author}
  {\bibfnamefont {M.}~\bibnamefont {Neeley}}, \bibinfo {author} {\bibfnamefont
  {A.}~\bibnamefont {Megrant}}, \bibinfo {author} {\bibfnamefont
  {B.}~\bibnamefont {Burkett}}, \bibinfo {author} {\bibfnamefont
  {R.}~\bibnamefont {Barends}}, \bibinfo {author} {\bibfnamefont
  {K.}~\bibnamefont {Arya}}, \bibinfo {author} {\bibfnamefont {B.}~\bibnamefont
  {Chiaro}}, \bibinfo {author} {\bibfnamefont {Y.}~\bibnamefont {Chen}},
  \bibinfo {author} {\bibfnamefont {A.}~\bibnamefont {Dunsworth}}, \bibinfo
  {author} {\bibfnamefont {A.}~\bibnamefont {Fowler}}, \bibinfo {author}
  {\bibfnamefont {B.}~\bibnamefont {Foxen}}, \bibinfo {author} {\bibfnamefont
  {C.}~\bibnamefont {Gidney}}, \bibinfo {author} {\bibfnamefont
  {M.}~\bibnamefont {Giustina}}, \bibinfo {author} {\bibfnamefont
  {R.}~\bibnamefont {Graff}}, \bibinfo {author} {\bibfnamefont
  {T.}~\bibnamefont {Huang}}, \bibinfo {author} {\bibfnamefont
  {E.}~\bibnamefont {Jeffrey}}, \bibinfo {author} {\bibfnamefont
  {E.}~\bibnamefont {Lucero}}, \bibinfo {author} {\bibfnamefont {J.~Y.}\
  \bibnamefont {Mutus}}, \bibinfo {author} {\bibfnamefont {O.}~\bibnamefont
  {Naaman}}, \bibinfo {author} {\bibfnamefont {C.}~\bibnamefont {Neill}},
  \bibinfo {author} {\bibfnamefont {C.}~\bibnamefont {Quintana}}, \bibinfo
  {author} {\bibfnamefont {P.}~\bibnamefont {Roushan}}, \bibinfo {author}
  {\bibfnamefont {D.}~\bibnamefont {Sank}}, \bibinfo {author} {\bibfnamefont
  {A.}~\bibnamefont {Vainsencher}}, \bibinfo {author} {\bibfnamefont
  {J.}~\bibnamefont {Wenner}}, \bibinfo {author} {\bibfnamefont {T.~C.}\
  \bibnamefont {White}}, \bibinfo {author} {\bibfnamefont {S.}~\bibnamefont
  {Boixo}}, \bibinfo {author} {\bibfnamefont {R.}~\bibnamefont {Babbush}},
  \bibinfo {author} {\bibfnamefont {V.~N.}\ \bibnamefont {Smelyanskiy}},
  \bibinfo {author} {\bibfnamefont {H.}~\bibnamefont {Neven}},\ and\ \bibinfo
  {author} {\bibfnamefont {J.~M.}\ \bibnamefont {Martinis}},\ }\href
  {https://doi.org/10.1103/PhysRevLett.121.090502} {\bibfield  {journal}
  {\bibinfo  {journal} {Physical Review Letters}\ }\textbf {\bibinfo {volume}
  {121}},\ \bibinfo {pages} {090502} (\bibinfo {year} {2018})}\BibitemShut
  {NoStop}\bibitem [{\citenamefont {Carroll}\ \emph {et~al.}(2022)\citenamefont
  {Carroll}, \citenamefont {Rosenblatt}, \citenamefont {Jurcevic},
  \citenamefont {Lauer},\ and\ \citenamefont {Kandala}}]{Carroll2022dynamics}\BibitemOpen
  \bibfield  {author} {\bibinfo {author} {\bibfnamefont {M.}~\bibnamefont
  {Carroll}}, \bibinfo {author} {\bibfnamefont {S.}~\bibnamefont {Rosenblatt}},
  \bibinfo {author} {\bibfnamefont {P.}~\bibnamefont {Jurcevic}}, \bibinfo
  {author} {\bibfnamefont {I.}~\bibnamefont {Lauer}},\ and\ \bibinfo {author}
  {\bibfnamefont {A.}~\bibnamefont {Kandala}},\ }\href
  {https://doi.org/10.1038/s41534-022-00643-y} {\bibfield  {journal} {\bibinfo
  {journal} {npj Quantum Information}\ }\textbf {\bibinfo {volume} {8}},\
  \bibinfo {pages} {132} (\bibinfo {year} {2022})},\ \Eprint
  {https://arxiv.org/abs/2105.15201} {2105.15201} \BibitemShut {NoStop}\bibitem [{\citenamefont {Kelly}\ \emph {et~al.}(2018)\citenamefont {Kelly},
  \citenamefont {O'Malley}, \citenamefont {Neeley}, \citenamefont {Neven},\
  and\ \citenamefont {Martinis}}]{Kelly2018physical}\BibitemOpen
  \bibfield  {author} {\bibinfo {author} {\bibfnamefont {J.}~\bibnamefont
  {Kelly}}, \bibinfo {author} {\bibfnamefont {P.}~\bibnamefont {O'Malley}},
  \bibinfo {author} {\bibfnamefont {M.}~\bibnamefont {Neeley}}, \bibinfo
  {author} {\bibfnamefont {H.}~\bibnamefont {Neven}},\ and\ \bibinfo {author}
  {\bibfnamefont {J.~M.}\ \bibnamefont {Martinis}},\ }\href@noop {} {\bibinfo
  {title} {Physical qubit calibration on a directed acyclic graph}},\ \bibinfo
  {howpublished} {arXiv:1803.03226} (\bibinfo {year} {2018}),\ \Eprint
  {https://arxiv.org/abs/1803.03226} {1803.03226} \BibitemShut {NoStop}\bibitem [{\citenamefont {Sarovar}\ \emph {et~al.}(2020)\citenamefont
  {Sarovar}, \citenamefont {Proctor}, \citenamefont {Rudinger}, \citenamefont
  {Young}, \citenamefont {Nielsen},\ and\ \citenamefont
  {Blume-Kohout}}]{Sarovar2020detectingcrosstalk}\BibitemOpen
  \bibfield  {author} {\bibinfo {author} {\bibfnamefont {M.}~\bibnamefont
  {Sarovar}}, \bibinfo {author} {\bibfnamefont {T.}~\bibnamefont {Proctor}},
  \bibinfo {author} {\bibfnamefont {K.}~\bibnamefont {Rudinger}}, \bibinfo
  {author} {\bibfnamefont {K.}~\bibnamefont {Young}}, \bibinfo {author}
  {\bibfnamefont {E.}~\bibnamefont {Nielsen}},\ and\ \bibinfo {author}
  {\bibfnamefont {R.}~\bibnamefont {Blume-Kohout}},\ }\href
  {https://doi.org/10.22331/q-2020-09-11-321} {\bibfield  {journal} {\bibinfo
  {journal} {{Quantum}}\ }\textbf {\bibinfo {volume} {4}},\ \bibinfo {pages}
  {321} (\bibinfo {year} {2020})}\BibitemShut {NoStop}\bibitem [{\citenamefont {Mundada}\ \emph {et~al.}(2019)\citenamefont
  {Mundada}, \citenamefont {Zhang}, \citenamefont {Hazard},\ and\ \citenamefont
  {Houck}}]{Mundada2019}\BibitemOpen
  \bibfield  {author} {\bibinfo {author} {\bibfnamefont {P.}~\bibnamefont
  {Mundada}}, \bibinfo {author} {\bibfnamefont {G.}~\bibnamefont {Zhang}},
  \bibinfo {author} {\bibfnamefont {T.}~\bibnamefont {Hazard}},\ and\ \bibinfo
  {author} {\bibfnamefont {A.}~\bibnamefont {Houck}},\ }\href
  {https://link.aps.org/doi/10.1103/PhysRevApplied.12.054023} {\bibfield
  {journal} {\bibinfo  {journal} {Phys. Rev. Applied}\ }\textbf {\bibinfo
  {volume} {12}},\ \bibinfo {pages} {054023} (\bibinfo {year}
  {2019})}\BibitemShut {NoStop}\bibitem [{\citenamefont {Ku}\ \emph {et~al.}(2020)\citenamefont {Ku},
  \citenamefont {Xu}, \citenamefont {Brink}, \citenamefont {McKay},
  \citenamefont {Hertzberg}, \citenamefont {Ansari},\ and\ \citenamefont
  {Plourde}}]{Ku2020}\BibitemOpen
  \bibfield  {author} {\bibinfo {author} {\bibfnamefont {J.}~\bibnamefont
  {Ku}}, \bibinfo {author} {\bibfnamefont {X.}~\bibnamefont {Xu}}, \bibinfo
  {author} {\bibfnamefont {M.}~\bibnamefont {Brink}}, \bibinfo {author}
  {\bibfnamefont {D.~C.}\ \bibnamefont {McKay}}, \bibinfo {author}
  {\bibfnamefont {J.~B.}\ \bibnamefont {Hertzberg}}, \bibinfo {author}
  {\bibfnamefont {M.~H.}\ \bibnamefont {Ansari}},\ and\ \bibinfo {author}
  {\bibfnamefont {B.~L.~T.}\ \bibnamefont {Plourde}},\ }\href
  {https://link.aps.org/doi/10.1103/PhysRevLett.125.200504} {\bibfield
  {journal} {\bibinfo  {journal} {Phys. Rev. Lett.}\ }\textbf {\bibinfo
  {volume} {125}},\ \bibinfo {pages} {200504} (\bibinfo {year}
  {2020})}\BibitemShut {NoStop}\bibitem [{\citenamefont {Kandala}\ \emph {et~al.}(2021)\citenamefont
  {Kandala}, \citenamefont {Wei}, \citenamefont {Srinivasan}, \citenamefont
  {Magesan}, \citenamefont {Carnevale}, \citenamefont {Keefe}, \citenamefont
  {Klaus}, \citenamefont {Dial},\ and\ \citenamefont
  {McKay}}]{Kandala2021demonstration}\BibitemOpen
  \bibfield  {author} {\bibinfo {author} {\bibfnamefont {A.}~\bibnamefont
  {Kandala}}, \bibinfo {author} {\bibfnamefont {K.~X.}\ \bibnamefont {Wei}},
  \bibinfo {author} {\bibfnamefont {S.}~\bibnamefont {Srinivasan}}, \bibinfo
  {author} {\bibfnamefont {E.}~\bibnamefont {Magesan}}, \bibinfo {author}
  {\bibfnamefont {S.}~\bibnamefont {Carnevale}}, \bibinfo {author}
  {\bibfnamefont {G.~A.}\ \bibnamefont {Keefe}}, \bibinfo {author}
  {\bibfnamefont {D.}~\bibnamefont {Klaus}}, \bibinfo {author} {\bibfnamefont
  {O.}~\bibnamefont {Dial}},\ and\ \bibinfo {author} {\bibfnamefont {D.~C.}\
  \bibnamefont {McKay}},\ }\href
  {https://doi.org/10.1103/PhysRevLett.127.130501} {\bibfield  {journal}
  {\bibinfo  {journal} {Phys. Rev. Lett.}\ }\textbf {\bibinfo {volume} {127}},\
  \bibinfo {pages} {130501} (\bibinfo {year} {2021})},\ \Eprint
  {https://arxiv.org/abs/2011.07050} {2011.07050} \BibitemShut {NoStop}\bibitem [{\citenamefont {Wittler}\ \emph {et~al.}(2021)\citenamefont
  {Wittler}, \citenamefont {Roy}, \citenamefont {Pack}, \citenamefont
  {Werninghaus}, \citenamefont {Roy}, \citenamefont {Egger}, \citenamefont
  {Filipp}, \citenamefont {Wilhelm},\ and\ \citenamefont
  {Machnes}}]{Wittler2021integrated}\BibitemOpen
  \bibfield  {author} {\bibinfo {author} {\bibfnamefont {N.}~\bibnamefont
  {Wittler}}, \bibinfo {author} {\bibfnamefont {F.}~\bibnamefont {Roy}},
  \bibinfo {author} {\bibfnamefont {K.}~\bibnamefont {Pack}}, \bibinfo {author}
  {\bibfnamefont {M.}~\bibnamefont {Werninghaus}}, \bibinfo {author}
  {\bibfnamefont {A.~S.}\ \bibnamefont {Roy}}, \bibinfo {author} {\bibfnamefont
  {D.~J.}\ \bibnamefont {Egger}}, \bibinfo {author} {\bibfnamefont
  {S.}~\bibnamefont {Filipp}}, \bibinfo {author} {\bibfnamefont {F.~K.}\
  \bibnamefont {Wilhelm}},\ and\ \bibinfo {author} {\bibfnamefont
  {S.}~\bibnamefont {Machnes}},\ }\href
  {https://doi.org/10.1103/PhysRevApplied.15.034080} {\bibfield  {journal}
  {\bibinfo  {journal} {Phys. Rev. Applied}\ }\textbf {\bibinfo {volume}
  {15}},\ \bibinfo {pages} {034080} (\bibinfo {year} {2021})},\ \Eprint
  {https://arxiv.org/abs/2009.09866} {2009.09866} \BibitemShut {NoStop}\bibitem [{\citenamefont {Shen}\ \emph {et~al.}(2026)\citenamefont {Shen},
  \citenamefont {Chen}, \citenamefont {Huang}, \citenamefont {Takeshita},
  \citenamefont {Vezvaee}, \citenamefont {Medalsy},\ and\ \citenamefont
  {Lidar}}]{shen2026qmctwinmasterequationsimulationsyndrome}\BibitemOpen
  \bibfield  {author} {\bibinfo {author} {\bibfnamefont {T.}~\bibnamefont
  {Shen}}, \bibinfo {author} {\bibfnamefont {H.}~\bibnamefont {Chen}}, \bibinfo
  {author} {\bibfnamefont {B.}~\bibnamefont {Huang}}, \bibinfo {author}
  {\bibfnamefont {T.}~\bibnamefont {Takeshita}}, \bibinfo {author}
  {\bibfnamefont {A.}~\bibnamefont {Vezvaee}}, \bibinfo {author} {\bibfnamefont
  {I.}~\bibnamefont {Medalsy}},\ and\ \bibinfo {author} {\bibfnamefont {D.~A.}\
  \bibnamefont {Lidar}},\ }\href {https://arxiv.org/abs/2606.19848} {\bibinfo
  {title} {{QMCtwin}: Master-equation simulation of syndrome statistics beyond
  {Pauli} noise}} (\bibinfo {year} {2026}),\ \Eprint
  {https://arxiv.org/abs/2606.19848} {arXiv:2606.19848 [quant-ph]} \BibitemShut
  {NoStop}\bibitem [{\citenamefont {Tripathi}\ \emph {et~al.}(2022)\citenamefont
  {Tripathi}, \citenamefont {Chen}, \citenamefont {Khezri}, \citenamefont
  {Yip}, \citenamefont {Levenson-Falk},\ and\ \citenamefont
  {Lidar}}]{tripathi2021suppression}\BibitemOpen
  \bibfield  {author} {\bibinfo {author} {\bibfnamefont {V.}~\bibnamefont
  {Tripathi}}, \bibinfo {author} {\bibfnamefont {H.}~\bibnamefont {Chen}},
  \bibinfo {author} {\bibfnamefont {M.}~\bibnamefont {Khezri}}, \bibinfo
  {author} {\bibfnamefont {K.-W.}\ \bibnamefont {Yip}}, \bibinfo {author}
  {\bibfnamefont {E.~M.}\ \bibnamefont {Levenson-Falk}},\ and\ \bibinfo
  {author} {\bibfnamefont {D.~A.}\ \bibnamefont {Lidar}},\ }\href
  {https://doi.org/10.1103/PhysRevApplied.18.024068} {\bibfield  {journal}
  {\bibinfo  {journal} {Physical Review Applied}\ }\textbf {\bibinfo {volume}
  {18}},\ \bibinfo {pages} {024068} (\bibinfo {year} {2022})}\BibitemShut
  {NoStop}\bibitem [{\citenamefont {Evert}\ \emph {et~al.}(2025)\citenamefont {Evert},
  \citenamefont {Gonzalez~Izquierdo}, \citenamefont {Sud}, \citenamefont {Hu},
  \citenamefont {Grabbe}, \citenamefont {Rieffel}, \citenamefont {Reagor},\
  and\ \citenamefont {Wang}}]{evert2024syncopated}\BibitemOpen
  \bibfield  {author} {\bibinfo {author} {\bibfnamefont {B.}~\bibnamefont
  {Evert}}, \bibinfo {author} {\bibfnamefont {Z.}~\bibnamefont
  {Gonzalez~Izquierdo}}, \bibinfo {author} {\bibfnamefont {J.}~\bibnamefont
  {Sud}}, \bibinfo {author} {\bibfnamefont {H.-Y.}\ \bibnamefont {Hu}},
  \bibinfo {author} {\bibfnamefont {S.}~\bibnamefont {Grabbe}}, \bibinfo
  {author} {\bibfnamefont {E.~G.}\ \bibnamefont {Rieffel}}, \bibinfo {author}
  {\bibfnamefont {M.~J.}\ \bibnamefont {Reagor}},\ and\ \bibinfo {author}
  {\bibfnamefont {Z.}~\bibnamefont {Wang}},\ }\href
  {https://doi.org/10.1103/8lxc-lvv1} {\bibfield  {journal} {\bibinfo
  {journal} {Physical Review Applied}\ }\textbf {\bibinfo {volume} {24}},\
  \bibinfo {pages} {044025} (\bibinfo {year} {2025})}\BibitemShut {NoStop}\bibitem [{\citenamefont {Lidar}\ and\ \citenamefont
  {Wu}(2002)}]{Wu2002recoupling}\BibitemOpen
  \bibfield  {author} {\bibinfo {author} {\bibfnamefont {D.~A.}\ \bibnamefont
  {Lidar}}\ and\ \bibinfo {author} {\bibfnamefont {L.-A.}\ \bibnamefont {Wu}},\
  }\href {https://doi.org/10.1103/PhysRevLett.88.017905} {\bibfield  {journal}
  {\bibinfo  {journal} {Phys. Rev. Lett.}\ }\textbf {\bibinfo {volume} {88}},\
  \bibinfo {pages} {017905} (\bibinfo {year} {2002})}\BibitemShut {NoStop}\bibitem [{\citenamefont {Huang}\ \emph {et~al.}(2023)\citenamefont {Huang},
  \citenamefont {Tong}, \citenamefont {Fang},\ and\ \citenamefont
  {Su}}]{Huang2023learning}\BibitemOpen
  \bibfield  {author} {\bibinfo {author} {\bibfnamefont {H.-Y.}\ \bibnamefont
  {Huang}}, \bibinfo {author} {\bibfnamefont {Y.}~\bibnamefont {Tong}},
  \bibinfo {author} {\bibfnamefont {D.}~\bibnamefont {Fang}},\ and\ \bibinfo
  {author} {\bibfnamefont {Y.}~\bibnamefont {Su}},\ }\href
  {https://doi.org/10.1103/PhysRevLett.130.200403} {\bibfield  {journal}
  {\bibinfo  {journal} {Phys. Rev. Lett.}\ }\textbf {\bibinfo {volume} {130}},\
  \bibinfo {pages} {200403} (\bibinfo {year} {2023})}\BibitemShut {NoStop}\bibitem [{\citenamefont {Gu}\ \emph {et~al.}(2024)\citenamefont {Gu},
  \citenamefont {Cincio},\ and\ \citenamefont {Coles}}]{Gu2024practical}\BibitemOpen
  \bibfield  {author} {\bibinfo {author} {\bibfnamefont {A.}~\bibnamefont
  {Gu}}, \bibinfo {author} {\bibfnamefont {L.}~\bibnamefont {Cincio}},\ and\
  \bibinfo {author} {\bibfnamefont {P.~J.}\ \bibnamefont {Coles}},\ }\href
  {https://doi.org/10.1038/s41467-023-44008-1} {\bibfield  {journal} {\bibinfo
  {journal} {Nat. Commun.}\ }\textbf {\bibinfo {volume} {15}},\ \bibinfo
  {pages} {312} (\bibinfo {year} {2024})}\BibitemShut {NoStop}\bibitem [{\citenamefont {Brown}\ and\ \citenamefont
  {Lidar}(2025)}]{brown2024efficient}\BibitemOpen
  \bibfield  {author} {\bibinfo {author} {\bibfnamefont {A.~F.}\ \bibnamefont
  {Brown}}\ and\ \bibinfo {author} {\bibfnamefont {D.~A.}\ \bibnamefont
  {Lidar}},\ }\href {https://doi.org/10.1103/1d4l-73x6} {\bibfield  {journal}
  {\bibinfo  {journal} {PRX Quantum}\ }\textbf {\bibinfo {volume} {6}},\
  \bibinfo {pages} {020354} (\bibinfo {year} {2025})}\BibitemShut {NoStop}\bibitem [{\citenamefont {Gebremedhin}\ \emph {et~al.}(2005)\citenamefont
  {Gebremedhin}, \citenamefont {Manne},\ and\ \citenamefont
  {Pothen}}]{Gebremedhin2005what}\BibitemOpen
  \bibfield  {author} {\bibinfo {author} {\bibfnamefont {A.~H.}\ \bibnamefont
  {Gebremedhin}}, \bibinfo {author} {\bibfnamefont {F.}~\bibnamefont {Manne}},\
  and\ \bibinfo {author} {\bibfnamefont {A.}~\bibnamefont {Pothen}},\ }\href
  {https://doi.org/10.1137/S0036144504444711} {\bibfield  {journal} {\bibinfo
  {journal} {SIAM Review}\ }\textbf {\bibinfo {volume} {47}},\ \bibinfo {pages}
  {629} (\bibinfo {year} {2005})}\BibitemShut {NoStop}\bibitem [{\citenamefont {Stollsteimer}\ and\ \citenamefont
  {Mahler}(2001)}]{Stollsteimer:01}\BibitemOpen
  \bibfield  {author} {\bibinfo {author} {\bibfnamefont {M.}~\bibnamefont
  {Stollsteimer}}\ and\ \bibinfo {author} {\bibfnamefont {G.}~\bibnamefont
  {Mahler}},\ }\href {https://doi.org/10.1103/PhysRevA.64.052301} {\bibfield
  {journal} {\bibinfo  {journal} {Physical Review A}\ }\textbf {\bibinfo
  {volume} {64}},\ \bibinfo {pages} {052301} (\bibinfo {year}
  {2001})}\BibitemShut {NoStop}\bibitem [{\citenamefont {Rotteler}\ and\ \citenamefont
  {Wocjan}(2006)}]{Rotteler:2006aa}\BibitemOpen
  \bibfield  {author} {\bibinfo {author} {\bibfnamefont {M.}~\bibnamefont
  {Rotteler}}\ and\ \bibinfo {author} {\bibfnamefont {P.}~\bibnamefont
  {Wocjan}},\ }\href {https://doi.org/10.1109/TIT.2006.880059} {\bibfield
  {journal} {\bibinfo  {journal} {IEEE Transactions on Information Theory}\
  }\textbf {\bibinfo {volume} {52}},\ \bibinfo {pages} {4171} (\bibinfo {year}
  {2006})}\BibitemShut {NoStop}\bibitem [{\citenamefont {Vezvaee}\ \emph {et~al.}(2025)\citenamefont
  {Vezvaee}, \citenamefont {Tripathi}, \citenamefont {Kowsari}, \citenamefont
  {Levenson-Falk},\ and\ \citenamefont
  {Lidar}}]{vezvaee2024virtualzgatessymmetric}\BibitemOpen
  \bibfield  {author} {\bibinfo {author} {\bibfnamefont {A.}~\bibnamefont
  {Vezvaee}}, \bibinfo {author} {\bibfnamefont {V.}~\bibnamefont {Tripathi}},
  \bibinfo {author} {\bibfnamefont {D.}~\bibnamefont {Kowsari}}, \bibinfo
  {author} {\bibfnamefont {E.}~\bibnamefont {Levenson-Falk}},\ and\ \bibinfo
  {author} {\bibfnamefont {D.~A.}\ \bibnamefont {Lidar}},\ }\href
  {https://doi.org/10.1103/PRXQuantum.6.020348} {\bibfield  {journal} {\bibinfo
   {journal} {PRX Quantum}\ }\textbf {\bibinfo {volume} {6}},\ \bibinfo {pages}
  {020348} (\bibinfo {year} {2025})}\BibitemShut {NoStop}\bibitem [{\citenamefont {Sundar}\ \emph {et~al.}(2026)\citenamefont {Sundar},
  \citenamefont {Evert}, \citenamefont {Geyko}, \citenamefont {Patterson},
  \citenamefont {Joseph},\ and\ \citenamefont {Shi}}]{Sundar2026plasma}\BibitemOpen
  \bibfield  {author} {\bibinfo {author} {\bibfnamefont {B.}~\bibnamefont
  {Sundar}}, \bibinfo {author} {\bibfnamefont {B.}~\bibnamefont {Evert}},
  \bibinfo {author} {\bibfnamefont {V.}~\bibnamefont {Geyko}}, \bibinfo
  {author} {\bibfnamefont {A.}~\bibnamefont {Patterson}}, \bibinfo {author}
  {\bibfnamefont {I.}~\bibnamefont {Joseph}},\ and\ \bibinfo {author}
  {\bibfnamefont {Y.}~\bibnamefont {Shi}},\ }\href
  {https://doi.org/10.1103/lxr6-t7vb} {\bibfield  {journal} {\bibinfo
  {journal} {Phys. Rev. Applied}\ }\textbf {\bibinfo {volume} {25}},\ \bibinfo
  {pages} {024077} (\bibinfo {year} {2026})}\BibitemShut {NoStop}\bibitem [{\citenamefont {Abrams}\ \emph {et~al.}(2020)\citenamefont {Abrams},
  \citenamefont {Didier}, \citenamefont {Johnson}, \citenamefont {da~Silva},\
  and\ \citenamefont {Ryan}}]{Abrams2020implementation}\BibitemOpen
  \bibfield  {author} {\bibinfo {author} {\bibfnamefont {D.~M.}\ \bibnamefont
  {Abrams}}, \bibinfo {author} {\bibfnamefont {N.}~\bibnamefont {Didier}},
  \bibinfo {author} {\bibfnamefont {B.~R.}\ \bibnamefont {Johnson}}, \bibinfo
  {author} {\bibfnamefont {M.~P.}\ \bibnamefont {da~Silva}},\ and\ \bibinfo
  {author} {\bibfnamefont {C.~A.}\ \bibnamefont {Ryan}},\ }\href
  {https://doi.org/10.1038/s41928-020-00498-1} {\bibfield  {journal} {\bibinfo
  {journal} {Nature Electronics}\ }\textbf {\bibinfo {volume} {3}},\ \bibinfo
  {pages} {744} (\bibinfo {year} {2020})}\BibitemShut {NoStop}\bibitem [{\citenamefont {Caldwell}\ \emph {et~al.}(2018)\citenamefont
  {Caldwell}, \citenamefont {Didier}, \citenamefont {Ryan}, \citenamefont
  {Sete}, \citenamefont {Hudson}, \citenamefont {Karalekas}, \citenamefont
  {Manenti}, \citenamefont {da~Silva}, \citenamefont {Sinclair} \emph
  {et~al.}}]{Caldwell2018parametrically}\BibitemOpen
  \bibfield  {author} {\bibinfo {author} {\bibfnamefont {S.~A.}\ \bibnamefont
  {Caldwell}}, \bibinfo {author} {\bibfnamefont {N.}~\bibnamefont {Didier}},
  \bibinfo {author} {\bibfnamefont {C.~A.}\ \bibnamefont {Ryan}}, \bibinfo
  {author} {\bibfnamefont {E.~A.}\ \bibnamefont {Sete}}, \bibinfo {author}
  {\bibfnamefont {A.}~\bibnamefont {Hudson}}, \bibinfo {author} {\bibfnamefont
  {P.}~\bibnamefont {Karalekas}}, \bibinfo {author} {\bibfnamefont
  {R.}~\bibnamefont {Manenti}}, \bibinfo {author} {\bibfnamefont {M.~P.}\
  \bibnamefont {da~Silva}}, \bibinfo {author} {\bibfnamefont {R.}~\bibnamefont
  {Sinclair}}, \emph {et~al.},\ }\href
  {https://doi.org/10.1103/PhysRevApplied.10.034050} {\bibfield  {journal}
  {\bibinfo  {journal} {Phys. Rev. Applied}\ }\textbf {\bibinfo {volume}
  {10}},\ \bibinfo {pages} {034050} (\bibinfo {year} {2018})}\BibitemShut
  {NoStop}\bibitem [{\citenamefont {Brown}\ and\ \citenamefont
  {Lidar}(2026)}]{brown2026graphspecdatarepo}\BibitemOpen
  \bibfield  {author} {\bibinfo {author} {\bibfnamefont {A.~F.}\ \bibnamefont
  {Brown}}\ and\ \bibinfo {author} {\bibfnamefont {D.~A.}\ \bibnamefont
  {Lidar}},\ }\href {https://doi.org/10.5281/zenodo.24671940} {\bibinfo {title}
  {{Data repository for ``GraphSpec: Whole-Chip Hamiltonian Spectroscopy by
  Graph Coloring''}}},\ \bibinfo {howpublished} {Zenodo,
  \url{https://doi.org/10.5281/zenodo.24671940}} (\bibinfo {year}
  {2026})\BibitemShut {NoStop}\end{thebibliography}

\end{document}